\documentclass[twocolumn]{aastex61}
\usepackage{graphicx,subfigure,amsmath, amsfonts, amssymb,footnote,color,epstopdf}
\usepackage[bookmarks=false]{hyperref}
\usepackage{apjfonts}
\hypersetup{
  colorlinks= true, %Colours links instead of ugly boxes
  urlcolor  = black, %Colour for external hyperlinks
  linkcolor = red, %Colour of internal links
  citecolor = blue %Colour of citations
} % makes stupid green boxes go away, paper now looks more like it will when published!
\submitjournal{ApJ}
\accepted{July 6, 2017}
\shorttitle{VLA and ALMA Imaging of NGC\,404}
\shortauthors{Kristina Nyland et al.}
\begin{document}

%%%%%%%%%%%%%%%%%%%%%%%%%%%%%%%%%%%%%%%%%%%%%%%
% TITLE AND AUTHOR
\title{A Multi-wavelength Study of the Turbulent Central Engine of the Low-mass AGN hosted by NGC\,404}
%\title{VLA and ALMA Imaging of the Turbulent Central Engine of the Low-mass AGN hosted by NGC\,404}
%\title{Resolved VLA and ALMA Imaging of the Low-mass AGN in the Center of NGC\,404}
%\title{The Turbulent Center of the Low-Mass AGN Host NGC\,404}
%\title{The Turbulent Center of the Low-Mass AGN Host NGC\,404 Revealed with Resolved VLA and ALMA Imaging}

\correspondingauthor{Kristina Nyland}
\email{knyland@nrao.edu}

\author{Kristina Nyland}
\affiliation{National Radio Astronomy Observatory, Charlottesville, VA 22903, USA}

\author{Timothy A. Davis}
\affiliation{School of Physics \&\ Astronomy, Cardiff University, Queens Buildings, The Parade, Cardiff, CF24 3AA, UK}

\author{Dieu D. Nguyen}
\affiliation{Department of Physics and Astronomy, University of Utah, 115 South 1400 East, Salt Lake City, Utah 84112, USA}

\author{Anil Seth}
\affiliation{Department of Physics and Astronomy, University of Utah, 115 South 1400 East, Salt Lake City, Utah 84112, USA}

\author{Joan M. Wrobel}
\affiliation{National Radio Astronomy Observatory, Socorro, NM 87801, USA}

\author{Atish Kamble}\altaffiliation{Black Hole Initiative, Harvard University, 20 Garden St. Cambridge, MA 02138, USA}
\affiliation{Harvard-Smithsonian Center for Astrophysics, 60 Garden St., Cambridge, MA, USA}

\author{Mark Lacy}
\affiliation{National Radio Astronomy Observatory, Charlottesville, VA 22903, USA}

\author{Katherine Alatalo}\altaffiliation{Hubble fellow}
\affiliation{The Observatories of the Carnegie Institution for Science, 813 Santa Barbara St., Pasadena, CA 91101, USA}

\author{Margarita Karovska}
\affiliation{Harvard-Smithsonian Center for Astrophysics, 60 Garden St., Cambridge, MA, USA}

\author{W. Peter Maksym}
\affiliation{Harvard-Smithsonian Center for Astrophysics, 60 Garden St., Cambridge, MA, USA}

\author{Dipanjan Mukherjee}
\affiliation{Research School of Astronomy and Astrophysics, Australian National University, Canberra, ACT 2611, Australia}

\author{Lisa M. Young}
\affiliation{Physics Department, New Mexico Institute of Mining and Technology, Socorro, NM 87801, USA}

%%%%%%%%%%%%%%%%%%%%%%%%%%%%%%%%%%%%%%%%%%%%%%%
% ABSTRACT
\begin{abstract}
The nearby dwarf galaxy NGC\,404 harbors a low-luminosity active galactic nucleus (AGN) powered by the lowest-mass (< 150,000 M$_{\odot}$) central massive black hole (MBH) with a dynamical mass constraint currently known, thus providing a rare low-redshift analog to the MBH ``seeds'' that formed in the early Universe.  Here, we present new imaging of the nucleus of NGC\,404 at 12--18~GHz with the Karl G. Jansky Very Large Array (VLA) and observations of the CO(2--1) line with the Atacama Large Millimeter/Submillimeter Array (ALMA).  For the first time, we have successfully resolved the nuclear radio emission, revealing a centrally peaked, extended source spanning 17~pc.  Combined with previous VLA observations, our new data place a tight constraint on the radio spectral index and indicate an optically-thin synchrotron origin for the emission.  The peak of the resolved radio source coincides with the dynamical center of NGC\,404, the center of a rotating disk of molecular gas, and the position of a compact, hard X-ray source.  We also present evidence for shocks in the NGC\,404 nucleus from archival narrowband {\it HST} imaging, {\it Chandra} X-ray data, and {\it Spitzer} mid-infrared spectroscopy, and discuss possible origins for the shock excitation.  Given the morphology, location, and steep spectral index of the resolved radio source, as well as constraints on nuclear star formation from the ALMA CO(2--1) data, we find the most likely scenario for the origin of the radio source in the center of NGC\,404 to be a radio outflow associated with a confined jet driven by the active nucleus.   
\end{abstract}

%%%%%%%%%%%%%%%%%%%%%%%%%%%%%%%%%%%%%%%%%%%%%%%
% SUBJECT HEADINGS
\keywords{galaxies: dwarf --- galaxies: active --- galaxies: nuclei -- galaxies: individual (NGC404) --- radio continuum: galaxies}

%%%%%%%%%%%%%%%%%%%%%%%%%%%%%%%%%%%%%%%%%%%%%%%
%%%%%%%%%%%%%%%%%%% INTRODUCTION %%%%%%%%%%%%%%%%%%%
%%%%%%%%%%%%%%%%%%%%%%%%%%%%%%%%%%%%%%%%%%%%%%%
\section{Introduction and Motivation}
Although our knowledge of galaxy evolution has made great strides over the past several decades, we still lack a complete understanding of the formation of the first galaxies.  A number of lines of evidence, such as the effects of feedback from active galactic nuclei (AGNs) on the host galaxies in which they reside, strongly suggest that the formation and growth of galaxies and massive black holes (MBHs) are inextricably linked \citep{kormendy+13, heckman+14}.  More specifically, AGN feedback is believed to contribute to the observed scaling relations between MBH and host galaxy properties, the regulation of cooling flows in clusters, galaxy-scale outflows, the suppression of SF, and the build-up of the red sequence of massive early-type galaxies \citep{croton+06, fabian+12, somerville+15, volonteri+16}.  Determining the evolutionary impact of AGN feedback over cosmic time is therefore of great importance.  

A key observational parameter in improving our understanding of the growth and evolution of MBHs and galaxies is the mass distribution of MBH ``seeds'' \citep{volonteri+10, vanwassenhove+10, greene+12, mezcua+17}.  The shape of this distribution at high redshift would help distinguish between different dominant MBH formation mechanisms (hierarchical merging vs. direct gas collapse), thus providing a strong constraint for cosmological simulations (e.g., \citealt{shirakata+16}).  Some progress has been made at high redshift with the detection of luminous quasars at $z>6$, which argues for heavier MBH seed masses of $\sim10^5$~M$_{\odot}$ and/or super-Eddington accretion (e.g., \citealt{mortlock+11, wu+15}).  However, high-redshift MBH mass estimates, along with detailed information on the properties of the host galaxy, are notoriously difficult to obtain.

Measurements of the properties and distribution of MBHs with $M_{\mathrm{BH}} < 10^6$~M$_{\odot}$ hosted by lower-mass galaxies at low redshifts may offer a more promising population for probing MBH seed formation observationally.  A number of recent studies have sought to characterize this population in a statistical sense (e.g., \citealt{reines+15, mezcua+16}), however, detailed studies of these objects capable of providing insights into MBH accretion and feedback physics are still rare, particularly at radio frequencies.  Detections of radio continuum emission associated with low-mass ($M_{\mathrm{BH}} < 10^6$~M$_{\odot}$) AGNs include NGC\,4395 \citep{ho+01, wrobel+01, wrobel+06}, GH10 \citep{greene+06, wrobel+08}, and NGC\,404 \citep{nyland+12}.

NGC\,404 is a nearby (D = $3.06$ $\pm$ $0.37$~Mpc; \citealt{karachentsev+02}), dwarf S0 galaxy harboring a candidate accreting MBH with $M_{\mathrm{BH}} < 1.5 \times 10^5$~M$_{\odot}$ ($3\sigma$ stellar dynamics upper limit; \citealt{nguyen+17}), making it the galaxy with the lowest-mass central MBH with a dynamical constraint known.  A number of lines of evidence suggest that the putative MBH in the center of NGC\,404 is actively accreting material.  NGC\,404 hosts the nearest low-ionization nuclear emission-line region (LINER) \citep{ho+97} believed to be powered in part by a low-luminosity active galactic nucleus (LLAGN; \citealt{binder+11}).  Additional evidence includes AGN-like nuclear mid-infrared emission line ratios \citep{satyapal+04}, UV and optical continuum variability on timescales of decades \citep{maoz+05, nguyen+17}, variable hot dust emission at near-infrared wavelengths \citep{seth+10}, the presence of an AGN power-law component in {\it HST} observations within the central 0.2$^{\prime \prime}$ \citep{nguyen+17}, and the identification of a hard X-ray nuclear point source characterized by a power-law spectrum \citep{binder+11}. 

Recent radio continuum observations also support the presence of an active nucleus.  \citet{nyland+12} detected a 5~GHz source in their sub-arcsecond-resolution ($\theta_{\mathrm{FWHM}} \approx 0.4^{\prime \prime}$) observations that is spatially-coincident with the location of the hard X-ray point source and the optical center of the galaxy, within the positional uncertainties.  Assuming a common origin, they argued that the faint radio and hard X-ray emission, with $\log(L_{5\,\mathrm{GHz}}/{\mathrm{W}}~{\mathrm{Hz}}^{-1}) = 17.88$ and $\log(L_{\mathrm{2-10\,\,keV}}/{\mathrm{erg~s}^{-1}}) = 37.08$, respectively, could conceivably be produced by an X-ray binary, nuclear star formation, a single supernova remnant, or an LLAGN powered by the central low-mass MBH.  They concluded that the most likely scenario for the origin of the radio and X-ray emission is an accreting MBH since the other explanations could not simultaneously explain the observed X-ray and radio emission, were inconsistent with the known conditions in the nucleus of NGC\,404, or were statistically unlikely.  

Follow-up very long baseline interferometric (VLBI) 1.5~GHz continuum observations of NGC\,404 with sub-parsec-scale spatial resolution were carried out in 2012 with the Very Long Baseline Array (Project 12B-164; P.I. K. Nyland) and the European VLBI Network (EVN; \citealt{paragi+14}).  No radio emission was detected in either dataset at their respective sensitivities of $\sim$16\,$\mu$Jy beam$^{-1}$ and $\sim$7\,$\mu$Jy beam$^{-1}$.  The brightness temperature limit implied by the EVN observations of \citet{paragi+14} is $T_{\mathrm{b}} <$ 2 $\times$ 10$^5$\,K, near the limit of $T_{\mathrm{b}}$ $\sim$ 10$^{5}$ K \citep{condon+92} distinguishing compact starbursts ($T_{\mathrm{b}}$ $<$ 10$^{5}$ K) from AGNs ($T_{\mathrm{b}}$ $>$ 10$^{5}$ K).  

%%%%%%%%%%%%%%%%%%%%%%%%%%%%%%%%%%%%%%%%%%%%%%%
\begin{deluxetable*}{ccccccccccc}[t!]
\tablecaption{Summary of VLA Observations of NGC\,404 \label{tab:vla}}
\tablecolumns{9}
\tablewidth{0pt}
\tablehead{
\colhead{$\nu$} & \colhead{Date} & \colhead{Array} & \colhead{rms} & \colhead{Beam} & \colhead{B.P.A.} & \colhead{$S_{\mathrm{peak}}$} & \colhead{$S_{\mathrm{int}}$} & \colhead{$\log(L)$}\\
\colhead{(GHz)} & \colhead{} & \colhead{} & \colhead{($\mu$Jy beam$^{-1}$)} & \colhead{($^{\prime \prime}$)} & \colhead{(degrees)} & \colhead{(mJy beam$^{-1}$)} & \colhead{(mJy)} & \colhead{(W Hz$^{-1}$)}\\
\colhead{(1)} & \colhead{(2)} & \colhead{(3)} & \colhead{(4)} & \colhead{(5)} & \colhead{(6)} & \colhead{(7)} & \colhead{(8)} & \colhead{(9)}}
\startdata
1.5    & 2011 April 11, 30 & B & 55 & 3.42 $\times$ 2.79 & 122 &  2.10 $\pm$ 0.08 & 2.83 $\pm$ 0.14 & 18.51 $\pm$ 0.02\\
5.0    & 2011 July 9 & A & 16  & 0.39 $\times$ 0.32 & 103 &  0.25 $\pm$ 0.01 & 0.66 $\pm$ 0.06 & 17.88 $\pm$ 0.04\\
7.5    & 2011 July 9 & A & 11 & 0.39 $\times$ 0.32 & 103 &  0.11 $\pm$ 0.01 & 0.47 $\pm$ 0.04 & 17.73 $\pm$ 0.04\\
15.0  & 2014 February 24, 26 & A & 1.5 & 0.15 $\times$ 0.11 & 81 &  0.04 $\pm$ 0.01 & 0.29 $\pm$ 0.03 & 17.53 $\pm$ 0.05\\
\enddata
\tablecomments{Column (1): Central observing frequency.  Column (2): Observing date.  Column (3): VLA array configuration.  Column (4): Image rms noise.  Column (5): Clean beam major $\times$ minor axis.  Column (6): Clean beam position angle.  Column (7): Peak intensity.  For the 1.5, 5.0, and 7.5~GHz data the peak intensity and error are measured using the JMFIT task in the Astronomical Image Processing System (AIPS) in which a fit to an elliptical Gaussian model is performed.  The errors include a 3\% uncertainty in the absolute flux density scale, added in quadrature.  For the 15~GHz data, the intensity is measured using the IMSTAT task in CASA and the error is calculated as $\sqrt{(N \times \sigma)^2+(0.03 \times S_{\mathrm{peak}})^2}$, where $N$ is the number of synthesized beams subtended by the source and $S_{\mathrm{peak}}$ is the peak intensity.  Column (8): Total integrated flux density.  The measurements and uncertainties were calculated in the same manner as described for the peak flux density in Column (7).  Column (9): Spectral luminosity.}
\end{deluxetable*}
%%%%%%%%%%%%%%%%%%%%%%%%%%%%%%%%%%%%%%%%%%%%%%%%

From a physical standpoint, the VLBA and EVN non-detections could be an indication that the emission detected previously in the center of NGC\,404 at lower spatial resolution in \citet{nyland+12} is not powered by MBH accretion.  Another possibility is that the radio source is characterized by extended emission in the form of a radio outflow on scales intermediate to those probed by the existing radio observations.  The fact that the detected 5~GHz emission was slightly extended ($\approx$ 8.4 $\times$ 5.3~pc) compared to the synthesized beam (5.9 $\times$ 4.8~pc) lends some support to this possibility.  This would also be consistent with the known radio-loud nature of NGC\,404 \citep{nyland+12}.  

The VLBA and EVN continuum upper limits in the NGC\,404 nucleus are precedented by observations of other low-mass AGNs in which radio sources were resolved-out when observed at milliarcsecond-scale spatial resolutions (though we note that due to the close proximity of NGC\,404, the physical scale probed is much smaller than in most AGN).  
An example is the million-solar-mass MBH in the center of the dwarf starburst galaxy Henize 2-10 \citep{reines+11, reines+12}.  Henize 2-10 was detected at $C$ (4--8~GHz) and $X$ (8--12~GHz) band at sub-arcsecond spatial resolution with the VLA, and is characterized by a classic core+jet radio AGN morphology.  However, follow-up VLBA observations of Henize 2-10 failed to detect the central radio core at milliarcsecond-scale spatial resolution.  Long Baseline Array observations with a spatial resolution intermediate to those of the previous radio studies proved successful \citep{reines+12} in detecting the compact nuclear radio source, emphasizing the importance of probing a wide range of spatial scales when studying weakly accreting MBHs in nearby dwarf galaxies.

Here, we report the results of new Karl G. Jansky Very Large Array (VLA) $Ku$-band (12-18~GHz) continuum observations of NGC\,404 that bridge the gap in spatial scales previously studied.  We also present the detection of circumnuclear CO(2--1) emission with the Atacama Large Millimeter/Submillimeter Array (ALMA) at near-matched spatial resolution to the new VLA data.  In Section~\ref{obs}, we describe our new VLA $Ku$-band and ALMA CO(2--1) observations and data reduction.  In Section~\ref{results}, we discuss the properties of the resolved radio and CO(2--1) emission in the nucleus of NGC\,404.  In Section~\ref{sec:origin}, we re-evaluate the origin of the resolved nuclear radio source.  We examine evidence for the presence of shocks in the center of NGC\,404 based on archival {\it HST}, {\it Chandra}, and {\it Spitzer} data, consider the possibility of localized AGN feedback, and discuss the implications of this work on our understanding of galaxy evolution in Section~\ref{sec:discussion}.  We summarize our results and suggest future areas of research in Section~\ref{conclu}.
  
%%%%%%%%%%%%%%%%%%%%%%%%%%%%%%%%%%%%%%%%%%%%%%%
%%%%%%%%%%%%%%%%% OBSERVATIONS %%%%%%%%%%%%%%%%%%%%%
%%%%%%%%%%%%%%%%%%%%%%%%%%%%%%%%%%%%%%%%%%%%%%%
\section{Data}
\label{obs}
\subsection{VLA}
NGC\,404 was observed with the VLA at $Ku$ band (12-18~GHz) in the A-configuration on February 24 and 26, 2014 (project ID: 14A-305).  The observations were split into two scheduling blocks (SBs) spanning 2.5 hours each for a total observing time of 5 hours (2.5 hours on-source).  We performed reference pointing scans at $X$-band using a standard correlator set-up three times during each SB.  We configured the Wideband Digital Architecture (WIDAR) correlator with 4 $\times$ 2048 MHz basebands centered at 13, 15, 16, and 17~GHz using the 3-bit samplers to cover the full 6~GHz bandwidth at $Ku$-band.  Each baseband was divided into 16 spectral windows (spws), each containing 64 channels.  The correlator integration time was set to one second.  

We performed phase referencing using the calibrator J0111+3906 every three minutes with a switching angle of 4$^{\circ}$.  The positional accuracy of our phase calibrator was $< 0.002^{\prime \prime}$.  We observed the standard flux density calibrator 3C48 to set the amplitude scale to an accuracy of 3\% and calibrate the bandpass \citep{perley+13}.  Pointing solutions were derived approximately once per hour at X-band following standard VLA high-frequency observing guidelines.  All flagging, calibration, and imaging was performed using the Common Astronomy Software Applications (CASA) package\footnote{http://casa.nrao.edu}  (version 4.1.0).  Antenna 21 experienced tracking malfunctions in both SBs and was thus flagged from our observations.  Each SB was independently calibrated using the CASA VLA calibration pipeline version 1.2.0\footnote{https://science.nrao.edu/facilities/vla/data-processing/pipeline}.  This pipeline performs all standard calibrations, runs the RFLAG algorithm to automatically flag radio frequency interference, and calculates data weights using the STATWT task.  

After combining our SBs into a single dataset, we formed images of the $Ku$-band data using the CLEAN task in CASA with an image size of $\approx$ 1$^{\prime}$ (2048 pixels) and a pixel size of 0.03$^{\prime \prime}$.  We utilized the MS-MFS algorithm in CLEAN, which performs a multi-scale, multi-frequency deconvolution \citep{rau+11}, by setting the parameters nterms = 2 and multiscale = [0, 5, 15].  We chose Briggs weighting \citep{briggs+95} with a robustness parameter of 0.5 to obtain the best compromise among sensitivity, spatial resolution, and sidelobe suppression.  We utilized the $w$-projection algorithm \citep{cornwell+05} by setting the parameters gridmode = `widefield' and wprojplanes = 128 to correct for the effects of non-coplanar baselines.  Our final image at a central frequency of 15~GHz has a synthesized beam size of $\theta_{\mathrm{FWHM}} = 0.15^{\prime \prime}$ and an rms noise of 1.5~$\mu$Jy~beam$^{-1}$.  

%%%%%%%%%%%%%%%%%%%%%%%%%%%%%%%%%%%%%%%%%%%%%%%
\subsection{ALMA}
ALMA observations in band 6 covering the CO(2--1) line at a rest frequency of 230~GHz were performed on October 31, 2015 (P.I. Anil Seth).  Full details on these data will be presented in Davis et al.\ (in preparation).  In brief, calibration was performed using the ALMA calibration pipeline.  No continuum sources were detected (see Figure~\ref{fig:radio_spectrum}).  Imaging was performed in CASA using the CLEAN task with Briggs weighting and a robustness parameter of 0.5.  We produced a continuum-subtracted spectral line cube using the UVCONTSUB task in CASA.  The final cube used in our analysis has a channel width of 10~km~s$^{-1}$, a sensitivity of 0.4~mJy~beam$^{-1}$ per channel, and a synthesized beam size of $0.06^{\prime \prime} \times 0.04^{\prime \prime}$ (0.9 $\times$ 0.6~pc).  

The ALMA observations also include continuum data centered at a frequency of about 230~GHz.  After combining the three continuum basebands spanning about 6~GHz with the line-free channels from the spectral line cube, we find a 5$\sigma$ upper limit of 51~$\mu$Jy~beam$^{-1}$ for the presence of a compact millimeter continuum source in the center of NGC\,404.  We include this upper limit in our radio spectral index plot shown in Figure~\ref{fig:radio_spectrum}.

%%%%%%%%%%%%%%%%%%%%%%%%%%%%%%%%%%%%%%%%%%%%%%%
%%%%%%%%%%%%%%%%%%% RESULTS %%%%%%%%%%%%%%%%%%%%%%%
%%%%%%%%%%%%%%%%%%%%%%%%%%%%%%%%%%%%%%%%%%%%%%%
\bigskip
\section{Results}
\label{results}

%%%%%%%%%%%%%%%%%%%%%%%%%%%%%%%%%%%%%%%%%%%%%%
\subsection{Radio Continuum Properties}

%%%%%%%%%%%%%%%%%%%%%%%%%%%%%%%%%%%%%%%%%%%%%%
\subsubsection{Morphology and Flux Density}
\begin{figure}
\centering
\includegraphics[clip=true, trim=0cm 0cm 3.25cm 2cm, width=3.4in]{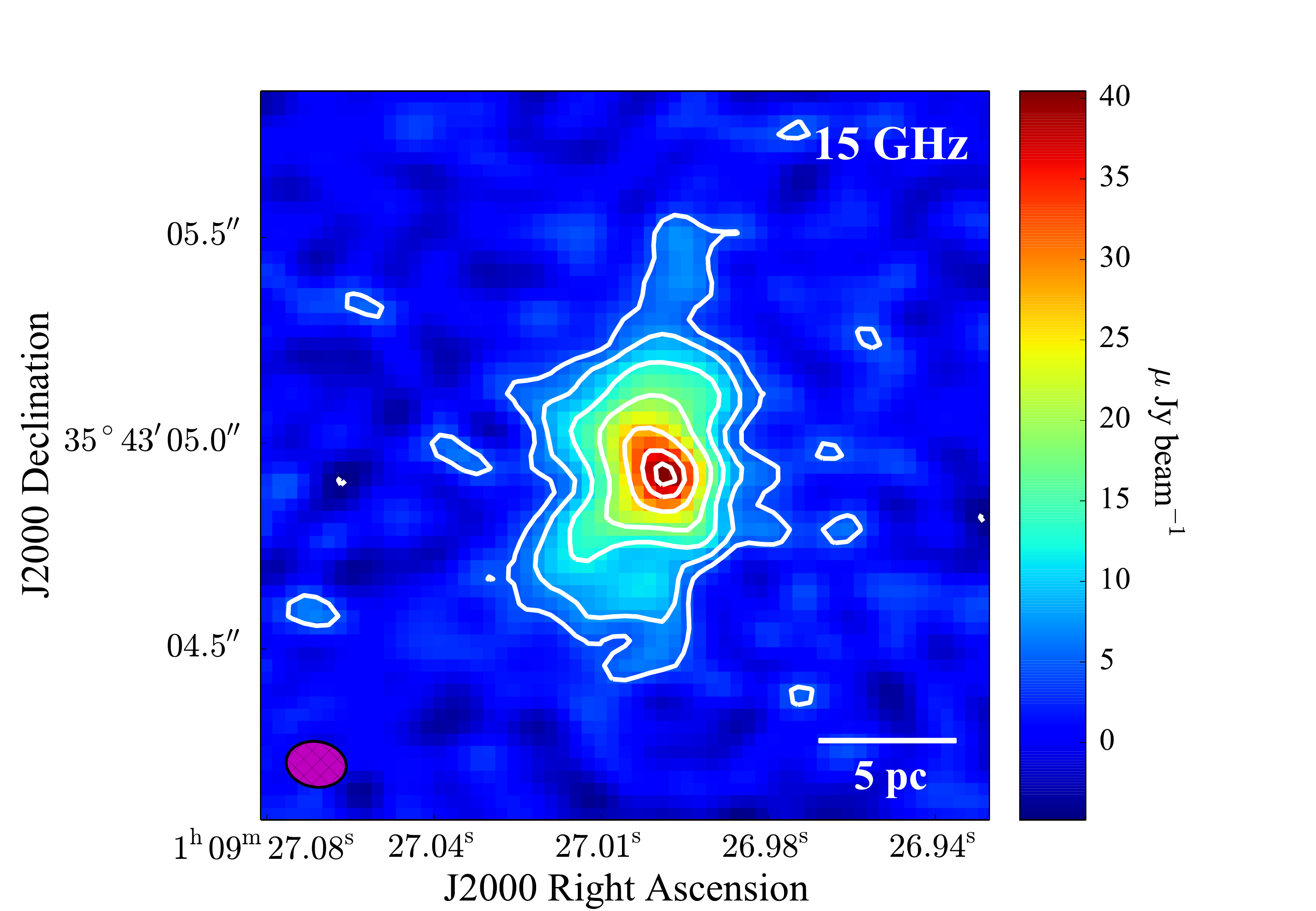} 
\caption{VLA continuum image over the frequency range 12$-$18~GHz of the nucleus of NGC\,404 with contours.  Negative contours are dashed.  The relative contour levels are [-3, 3, 5, 8, 12, 18, 23, 26] and the unit contour level represents the rms noise of 1.5 $\mu$Jy beam$^{-1}$.  The synthesized beam has major $\times$ minor axis dimensions of 0.15$^{\prime \prime} \times$ 0.11$^{\prime \prime}$ and is shown in the lower-left corner of the figure as a hatched magenta ellipse.  A scale bar denoting 5~pc (0.3$^{\prime \prime}$) is shown on the lower right.}
\label{fig:radio_overlay}
\end{figure}

We summarize the basic observational properties of our new $Ku$-band data, as well as the $L$- and $C$-band data presented in \citet{nyland+12}, in Table~\ref{tab:vla}.
A map of the 15~GHz emission with contours is shown in Figure~\ref{fig:radio_overlay}.  The full-width at half maximum (FWHM) of the major axis of the synthesized beam in this map is $\theta_{\mathrm{FWHM}}$ $\approx$ 0.15$\arcsec$ ($\sim$2.25~pc).  The spatially-resolved 15~GHz source has a major axis extent of 1.13$^{\prime \prime} \approx$ 17~pc with a position angle of 169.3 $\pm$ 1.3 degrees\footnote{Due to the complex nature of the resolved radio continuum source, we caution readers that this position angle measurement may carry a significantly larger uncertainty.}.  The integrated flux density is 0.29 $\pm$ 0.03~mJy, with a peak intensity of 0.04 $\pm$ 0.01~mJy~beam$^{-1}$ (see Table~\ref{tab:vla}).

%%%%%%%%%%%%%%%%%%%%%%%%%%%%%%%%%%%%%%%%%%%%%%%
%%%%%%%%%%%%%%%%%%%%%%%%%%%%%%%%%%%%%%%%%%%%%%%

%%%%%%%%%%%%%%%%%%%%%%%%%%%%%%%%%%%%%%%%%%%%%%%
\begin{figure}
\centering
\includegraphics[clip=true, trim=0cm 0cm 2cm 2cm, width=3.4in]{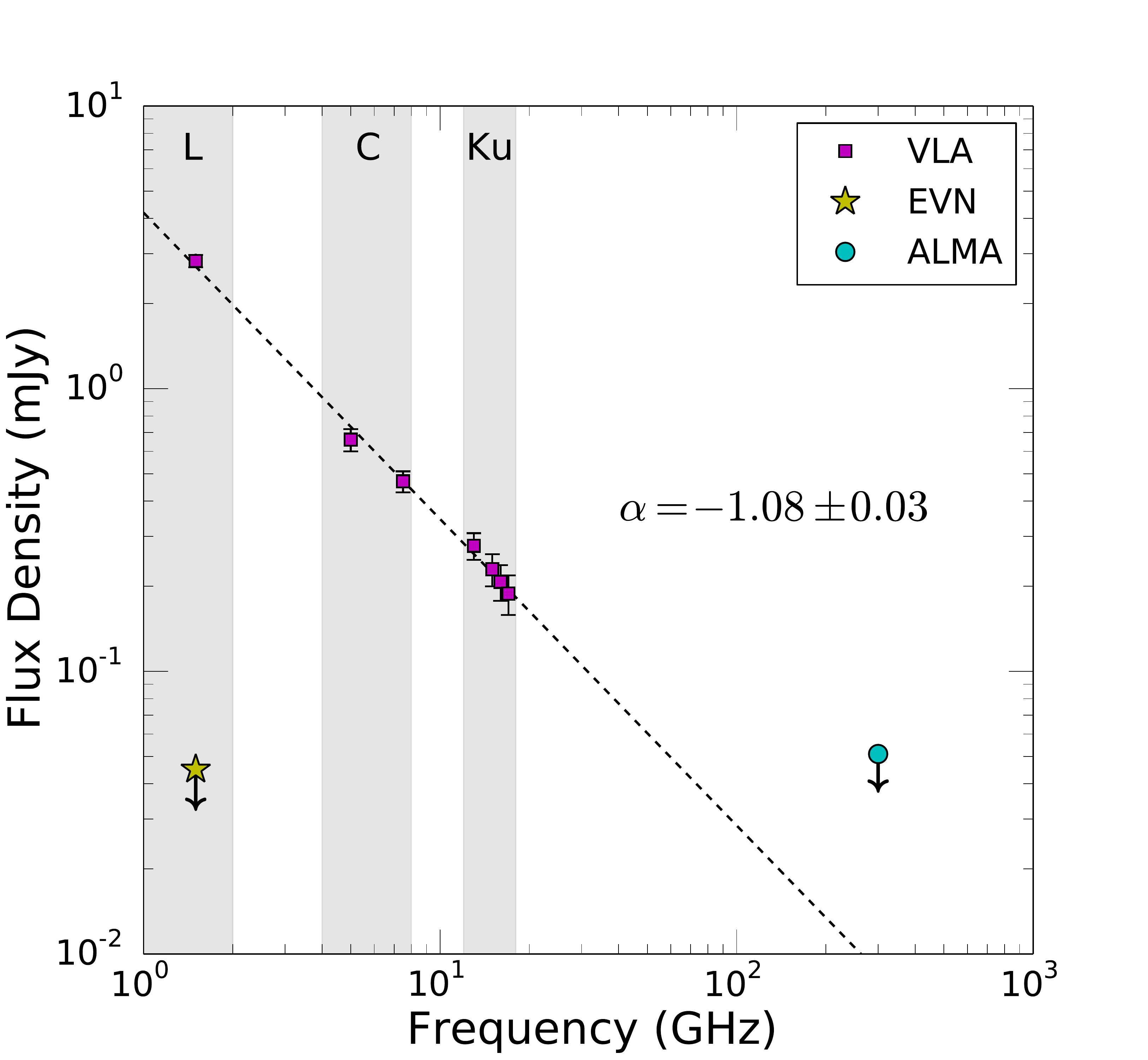} 
\caption{Radio spectrum of NGC\,404.  The magenta points are from VLA observations in the VLA $L$, $C$, and $Ku$ bands (\citealt{nyland+12}, this paper).  The dashed blue line shows the fitted spectral index of the VLA data of $\alpha = -1.08 \pm 0.03$.  The yellow star represents the upper limit to the presence of a sub-parsec-scale nuclear radio source from EVN observations \citep{paragi+14}.  The cyan point denotes the 230~GHz continuum upper limit from the continuum data associated with our ALMA band 6 observations.}
\label{fig:radio_spectrum}
\end{figure}
%%%%%%%%%%%%%%%%%%%%%%%%%%%%%%%%%%%%%%%%%%%%%%%

%%%%%%%%%%%%%%%%%%%%%%%%%%%%%%%%%%%%%%%%%%%%%%
\subsubsection{Radio Spectral Index}
In Figure~\ref{fig:radio_spectrum}, we show the radio spectrum of NGC\,404 from 1 to 18~GHz.  After tapering the $Ku$-band data to match the angular resolution of the $C$-band data presented in \citet{nyland+12}, we compute the radio spectral index over this range - which spans more than a factor of ten in frequency - of $\alpha = -1.08$ $\pm$ 0.03.   The 1$-$18~GHz radio spectral index is consistent with the in-band 4$-$8~GHz and the 1.5$-$8~GHz spectral index measurements reported in \citet{nyland+12} of $\alpha = -0.80 \pm 0.30$ and $\alpha = -1.15$ $\pm$ 0.05, respectively.  The steep spectral index is incompatible with pure thermal emission and suggests an optically-thin, non-thermal origin for the radio source \citep{condon+92}.  

We also include the upper limit to the 230~GHz (1~mm) continuum emission from the ALMA band 6 observations in Figure~\ref{fig:radio_spectrum}.  Millimeter continuum emission may arise from thermal processes (free-free emission or the Rayleigh-Jeans tail of thermal dust emission heated by young stars or an active nucleus), non-thermal processes (synchrotron emission associated with a radio jet), or a combination of these mechanisms \citep{honig+08, krips+11}.  Distinguishing between these possibilities, and constraining the relative contribution of each emission mechanism, would require spectral energy distribution (SED) modeling of additional (sub-)millimeter and infrared data at sub-arcsecond resolution (e.g., following the strategy of \citealt{koay+16}).  However, sufficient measurements at these wavelengths (particularly in the mid-infrared) with high enough resolution to isolate the nuclear component of the continuum emission are currently lacking for NGC\,404.

%%%%%%%%%%%%%%%%%%%%%%%%%%%%%%%%%%%%%%%%%%%%%%%
\begin{figure*}
\includegraphics[clip=true, trim=0cm 0.25cm 0cm 0cm, width=3.5in]{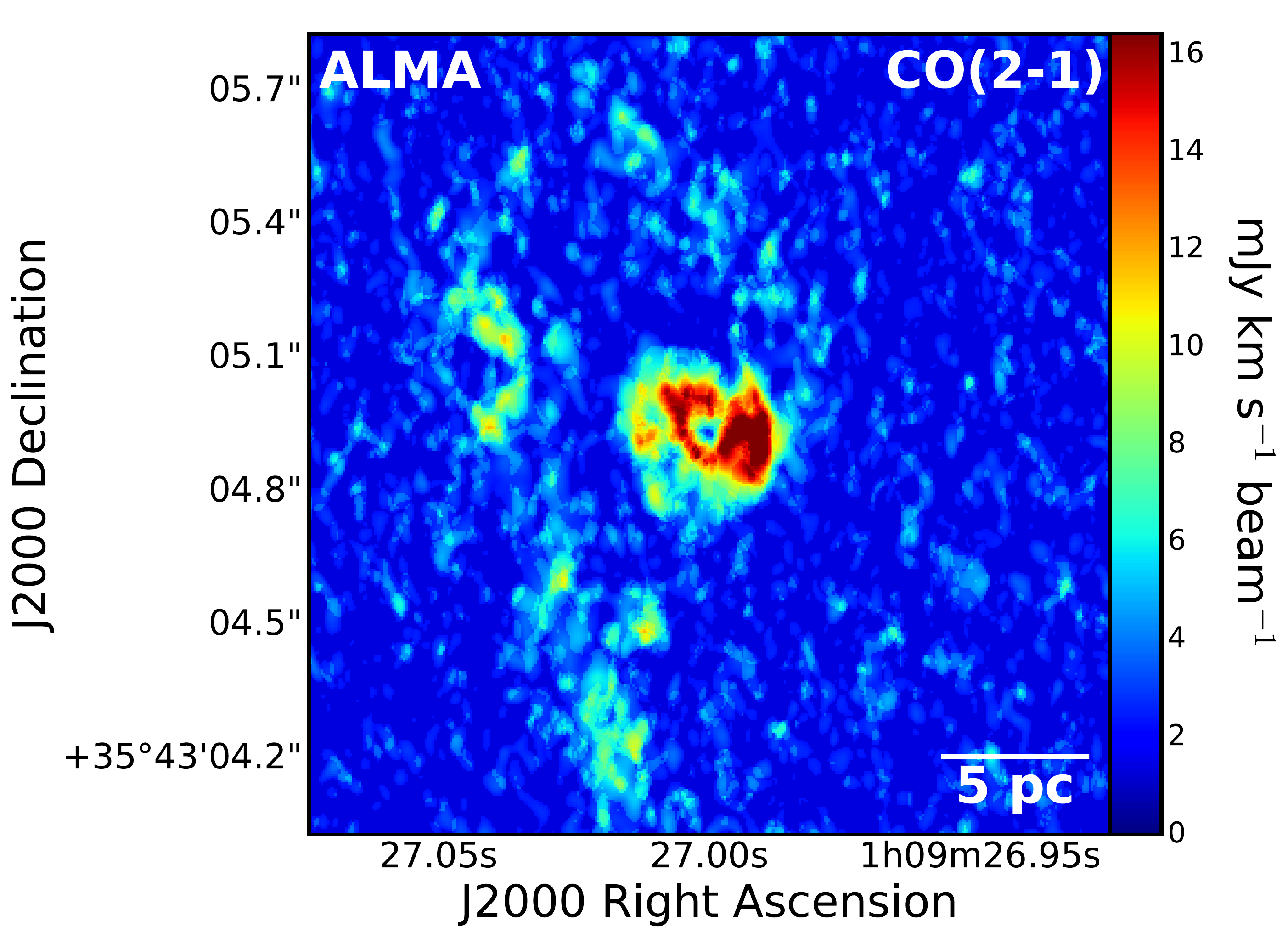} 
\includegraphics[clip=true, trim=0cm 0.25cm 0.25cm 0cm, width=3.5in]{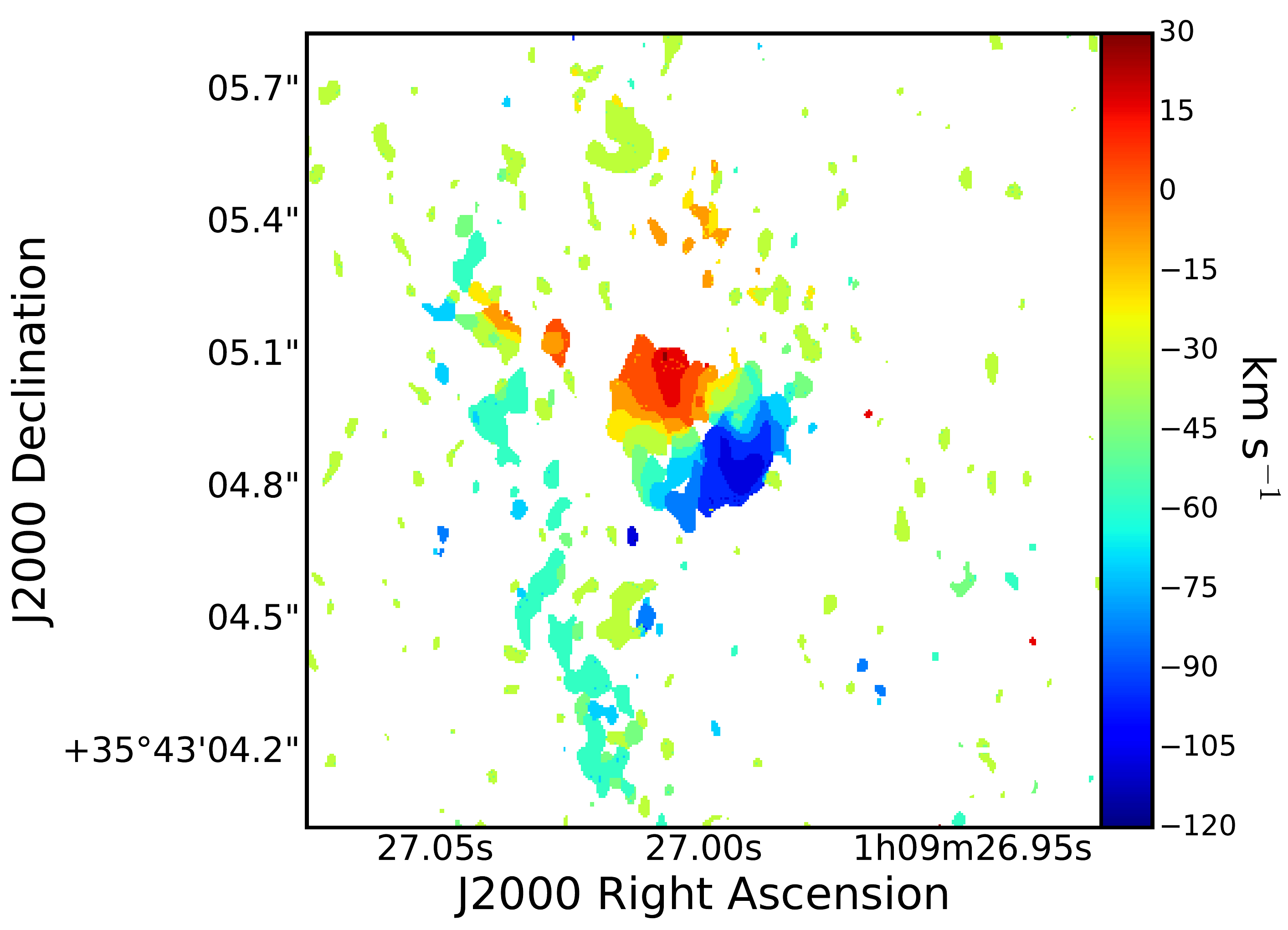} 
\caption{Moment 0 and 1 maps of the center of NGC\,404 from the ALMA spectral line data, each with dimensions of 1.8$^{\prime \prime} \times$ 1.8$^{\prime \prime}$ per side.  {\bf Left:} CO(2--1) integrated intensity (moment 0) map.  A scale bar denoting 5~pc (0.3$^{\prime \prime}$) is shown on the lower right.  {\bf Right:} CO(2--1) velocity (moment 1) map.  The systemic velocity of NGC\,404 is at $-48$~km~s$^{-1}$.  
%The ALMA synthesized beam is $0.06^{\prime \prime} \times 0.04^{\prime \prime}$ (0.9 $\times$ 0.6 pc).
\\}
\label{fig:alma}
\end{figure*}
%%%%%%%%%%%%%%%%%%%%%%%%%%%%%%%%%%%%%%%%%%%%%%%

%%%%%%%%%%%%%%%%%%%%%%%%%%%%%%%%%%%%%%%%%%%%%%
\medskip
\subsubsection{Radio Energetics}
\label{sec:radio_energetics}
The previous study of the nuclear radio emission associated with NGC\,404 presented in \citet{nyland+12} lacked sufficient spatial resolution to accurately measure the extent of the radio source.  The robust radio source size measurement presented here allows us to calculate the minimum synchrotron energy and magnetic field strength assuming equipartition between the particle and magnetic energy densities.  We assume a prolate spheroid geometry ($V = \frac{\pi}{6} \times d_{\mathrm{maj}} \times d_{\mathrm{min}}^2$) as in \citet{mezcua+15} with $d_{\mathrm{maj}} = 1.13^{\prime \prime}$ and $d_{\mathrm{min}} = 0.57^{\prime \prime}$.  We use Equations~\ref{eq:Emin} and \ref{eq:Bmin} below (based on \citealt{moffet+75}) to determine the minimum energy ($E_{\mathrm{min}}$) and magnetic field strength ($B_{\mathrm{min}}$) of the radio source:

\begin{equation}
\label{eq:Emin}
E_{\mathrm{min}} = 0.5(aAL)^{4/7}V^{3/7}
\end{equation}

\begin{equation}
\label{eq:Bmin}
B_{\mathrm{min}}=2.3(aAL/V)^{2/7},
\end{equation}
where $a$ is the contribution of protons relative to electrons to the total particle energy, $L$ is the luminosity of the source, and $V$ is the volume of the synchrotron emitting region.  For $a$, we assume the Milky Way value of $a = 100$ (e.g., \citealt{beck+01}), however, we note that this assumption carries a large uncertainty since  measurements of $a$ in external galaxies are not currently available.  The term $A$ in Equations~\ref{eq:Emin} and \ref{eq:Bmin} is defined as follows:

\begin{equation}
 A = C\, \frac{2\alpha + 2}{2\alpha + 1} \frac{\nu_{2}^{\alpha + 1/2} - \nu_{1}^{\alpha + 1/2}}{\nu_{2}^{\alpha + 1} - \nu_{1}^{\alpha + 1}},
 \end{equation}
where $C$ is a constant\footnote{The cgs units of $C$ are (g/cm)$^{3/4}$ s$^{-1}$.} of value 1.057 $\times$ 10$^{12}$, $\alpha$ is the spectral index, and $\nu_1$ and $\nu_2$ are the assumed lower and upper frequencies of the radio spectrum, respectively.  We assume $\alpha = -1.08$, $\nu_1 = 10$~MHz, and $\nu_2 = 100$~GHz.  We find a minimum energy of $E_{\mathrm{min}} \approx 1.54 \times 10^{50}$ erg and magnetic field strength of $B_{\mathrm{min}} \approx 292$~$\mu$G.

%%%%%%%%%%%%%%%%%%%%%%%%%%%%%%%%%%%%%%%%%%%%%%
\subsection{Molecular Gas Properties}
The CO(2--1) moment 0 (integrated intensity) map shown in Figure~\ref{fig:alma} is dominated by a bright, resolved, molecular disk with a central hole and a total extent of $0.4^{\prime \prime} \approx 6.5$~pc.  This source is located at $\alpha_{\mathrm{J2000}}$~=~01$^{\mathrm{h}}$09$^{\mathrm{m}}$27.000$^{\mathrm{s}}$ and $\delta_{\mathrm{J2000}}$~=~+35$\degr$43$\arcmin$04.93$\arcsec$, spatially coincident with the position of the peak of the radio continuum emission.  The molecular gas mass may be estimated using the following equation from \citet{bolatto+13}:

\begin{equation}
\label{eq:mol_gas_mass}
M_{\mathrm{mol}} = 1.05 \times 10^4  \left( \frac{X_{\mathrm{CO}}}{2\times10^{20} \frac{\mathrm{cm}^{-2}}{\mathrm{K}\,\mathrm{km}\,\mathrm{s}^{-1}}}\right) \frac{S_{\mathrm{CO}} \Delta v D^2}{(1+z)}  \,\,\ \mathrm{M}_{\odot},
\end{equation}
where $X_{\mathrm{CO}}$ is the CO-to-H$_2$ conversion factor, $S_{\mathrm{CO}} \Delta v$ is the integrated CO(1--0) line flux in units of Jy~km~s$^{-1}$, $D$ is the distance in units of Mpc, and $z$ is the redshift of the source.  

For NGC\,404, $z = -0.00016$.  We assume a standard Milky Way $X_{\mathrm{CO}}$ value of 2.3 $\times$ 10$^{20}$ cm$^{-2}$~(K~km~s$^{-1}$)$^{-1}$ \citep{bolatto+13} and a CO(2--1)/CO(1--0) flux density ratio of 3.2 \citep{carilli+13}.  Using Equation~\ref{eq:mol_gas_mass} with the integrated CO(2--1) line flux of 2.68~Jy~km~s$^{-1}$ (measured over a linewidth of $\sim$100~km~s$^{-1}$) and accounting for the CO(2--1)/CO(1--0) ratio\footnote{For completeness, we note that the average Milky Way value of the line ratio CO(2--1)/CO(1--0) = 3.2 that we use in our analysis is equivalent to $T_{\mathrm{B}}$(2--1)/$T_{\mathrm{B}}$(1--0) = 0.8.}, we obtain a molecular gas mass of $M_{\mathrm{H}_2} = 1.1 \times 10^5$~M$_{\odot}$ for the nuclear CO(2--1) disk.  This mass, which is similar to typical giant molecular cloud (GMC) masses in the Milky Way \citep{solomon+87}, is about 1\% of the total molecular gas mass measured in \citet{taylor+15} of 9.0 $\times$ 10$^6$ M$_{\odot}$ (or 7.7 $\times$ 10$^6$~M$_{\odot}$ after converting to the distance of 3.06~Mpc adopted in our study) at a spatial resolution of 7.6$^{\prime \prime}$ using the Berkeley-Illinois-Maryland Association (BIMA) interferometer.  This is not unexpected given the substantial difference in synthesized beam size between the BIMA and ALMA CO observations.  Thus, the ALMA CO(2--1) observations have likely resolved out most of the molecular gas detected in \citet{taylor+15}.  

The moment 0 map also reveals a more extended, curved feature to the west of the central brighter source spanning 1.7$^{\prime \prime} \approx 25.8$~pc. This CO(2--1) structure has a slightly higher integrated line flux compared to the central disk of 4.03~Jy~km~s$^{-1}$, which corresponds to a molecular gas mass of $M_{\mathrm{H}2} = 1.7 \times 10^5$~M$_{\odot}$.  Possible origins for the extended, low-surface-brightness molecular gas structure include a faint nuclear spiral arm or a tidally-disrupted GMC that is feeding cold gas onto the central MBH.  Additional ALMA observations with higher surface brightness sensitivity will ultimately be needed to distinguish between these possibilities.

In Figure~\ref{fig:alma}, we also show the CO(2--1) moment 1 map, which is characterized by a velocity field consistent with a regularly rotating gas disk on scales of a few parsecs.  The range of velocities spans about $-120$ to $30$~km~s$^{-1}$, which we note is significantly larger than the range velocities over which \citet{taylor+15} detected CO(1-0) emission in their BIMA observations.  This discrepancy is likely due to the substantially higher sensitivity of our ALMA observations.  Additional information on the properties of the CO(2--1) data, along with new MBH constraints from dynamical modeling using the ALMA data, will appear in Davis et al.\ (in preparation).  

%%%%%%%%%%%%%%%%%%%%%%%%%%%%%%%%%%%%%%%%%%%%%%%
\begin{figure*}
\centering
\includegraphics[clip=true, trim=0cm 8.25cm 0cm 6.5cm, width=7in]{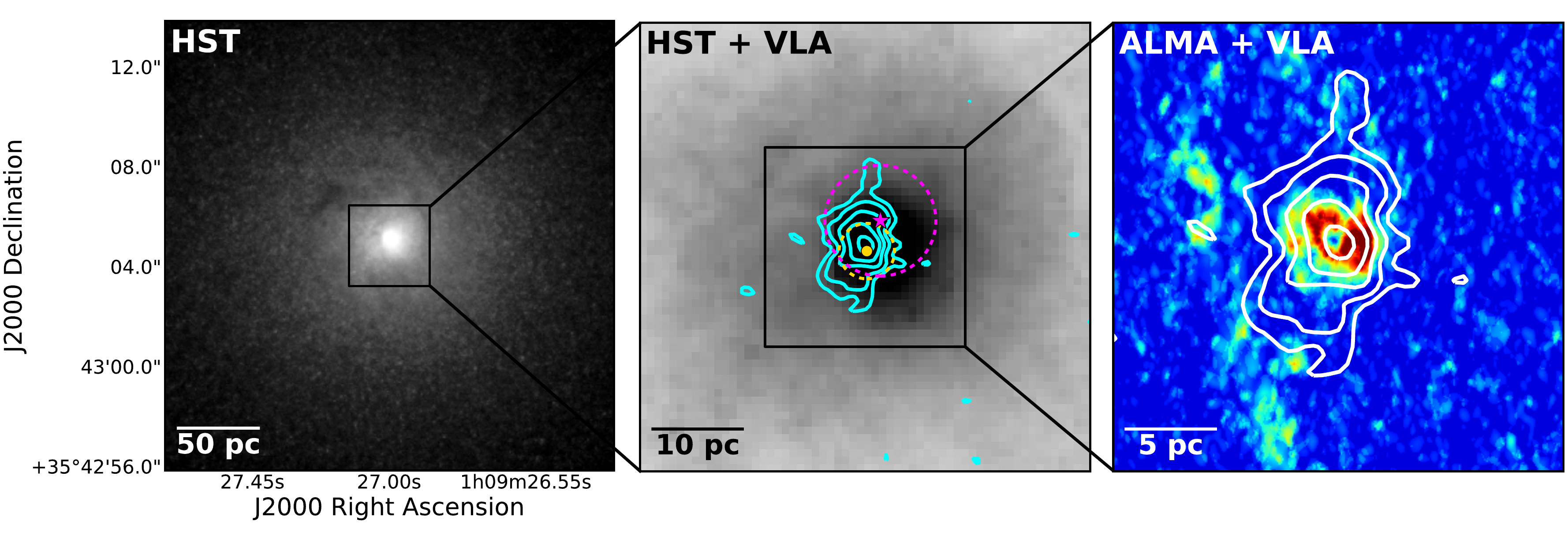} 
\caption{{\bf Left:} HST WFPC2 F814W image of the nucleus of NGC404 from the Hubble Legacy Archive.  The image measures 18.0$^{\prime \prime}$ (270.5~pc) per side.  {\bf Center:} Zoomed-in version of the HST image shown in the left panel with a length of 3.2$^{\prime \prime}$ (48.7~pc) per side.  The VLA 15~GHz contours are overlaid in cyan and are defined as in Figure~\ref{fig:radio_overlay}.  The magenta star marks the position of the hard (2-10 keV) X-ray point source and the dashed magenta circle shows the {\it Chandra} positional uncertainty of 0.4$^{\prime \prime}$ (6~pc) from \citet{binder+11}.  The yellow circle marks the location of the kinematic center of the NGC\,404 nucleus \citep{nguyen+17}.  The dashed yellow circle shows the positional uncertainty of the {\it HST} image of 0.2$^{\prime \prime}$.  
{\bf Right:} Zoomed-in view of the nucleus of NGC\,404 spanning 1.6$^{\prime \prime}$ (24~pc) in length per side.  The background colorscale shows the CO(2--1) integrated intensity map from our ALMA band 6 observations and the contours trace the VLA 15~GHz emission in white.
\\}
\label{fig:radio_alma_hst}
\end{figure*}
%%%%%%%%%%%%%%%%%%%%%%%%%%%%%%%%%%%%%%%%%%%%%%%

%%%%%%%%%%%%%%%%%%%%%%%%%%%%%%%%%%%%%%%%%%%%%%
\subsection{Astrometry}
\label{sec:astrometry}
The position of the peak of the 15~GHz emission is located at $\alpha_{\mathrm{J2000}}$~=~01$^{\mathrm{h}}$09$^{\mathrm{m}}$26.999$^{\mathrm{s}}$ and $\delta_{\mathrm{J2000}}$~=~+35$\degr$43$\arcmin$04.91$\arcsec$, identical to the radio position at 5~GHz reported in \citet{nyland+12}.  The conservative one-dimensional positional uncertainty of this measurement is 0.1$^{\prime \prime}$, and is dominated by our calibration strategy and the positional uncertainty of our phase-reference calibrator.  The VLA position is well-matched to that of the center of rotation of the CO(2--1) nuclear disk detected by ALMA, with a spatial offset of only 0.02$^{\prime \prime}$.  In Figure~\ref{fig:radio_alma_hst}, we show an optical {\it HST} image of NGC\,404.  As described in \citet{nguyen+17}, these {\it HST} data have a positional uncertainty of 0.2$^{\prime \prime}$.  Thus, within the known positional uncertainties, Figure~\ref{fig:radio_alma_hst} demonstrates that the radio continuum and central rotating CO(2--1) emission both appear to spatially coincident with the center of NGC\,404.  We note that we use the central position based on the stellar kinematics; this differs by 0.025$^{\prime \prime}$ (0.37~pc) from the hot-dust-corrected photocenter \citep{seth+10, nguyen+17}.  

The central region of NGC\,404 also hosts a hard X-ray source with a power law spectrum that has previously been studied in detail at X-ray wavelengths \citep{binder+11, paragi+14}.  Figure~\ref{fig:radio_alma_hst} shows the location of the hard X-ray source with respect to the optical and radio emission, along with the {\it Chandra} positional uncertainty reported in \citet{binder+11} of 0.4$^{\prime \prime}$.  The spatial coincidence between the radio, X-ray, and optical emission was a key factor in the conclusion of \citet{nyland+12} that the compact radio source detected at 5~GHz was most likely associated with an accreting low-mass AGN.  This highlights the importance of {\it Chandra} X-ray data for improving our understanding of low-mass AGNs (e.g., \citealt{reines+16}).

The astrometry of the NGC\,404 {\it Chandra} data was recently re-assessed by \citet{paragi+14}.  This study, which presented an analysis of new {\it Chandra} observations of NGC\,404, found a positional offset of 1.1$^{\prime \prime}$ with respect to previous {\it Chandra} observations, possibly due to an issue with the aspect reconstruction of the data used in one or both of the studies.  However, since the time of publication, newer versions of the data from \citet{binder+11} and \citet{paragi+14} are now available on the {\it Chandra} archive that have been reprocessed using improved calibration heuristics.  Comparing the reprocessed data from \citet{paragi+14} with the deeper observations from \citet{binder+11}, we find an offset of about 0.4$^{\prime \prime}$, which is consistent with the typical astrometric accuracy for {\it Chandra}.  Given the currently available data, we conclude as in \citet{nyland+12} that, within the known positional uncertainties, the hard X-ray source is most likely spatially coincident with the peak of the radio continuum source and the optical center of the galaxy.

%%%%%%%%%%%%%%%%%%%%%%%%%%%%%%%%%%%%%%%%%%%%%%%
%%%%%%%%%%%%%%%%%%%%%%%%%%%%%%%%%%%%%%%%%%%%%%%
%%%%%%%%%%%%%%%%%%%%%%%%%%%%%%%%%%%%%%%%%%%%%%%
\section{Origin of the Extended Radio Source}
\label{sec:origin}
Due to the low luminosity and diffuse morphology of the radio continuum emission in the center of NGC\,404, there are multiple possibilities for the physical origin of the radio source detected in our new 15~GHz VLA observations.  \citet{nyland+12} investigated a number of scenarios for the origin of the compact 5~GHz emission detected at lower spatial resolution, concluding that the accreting black hole scenario was the most likely explanation.  However, the possibility that the radio source is instead associated with star formation or a single ultraluminous radio supernova remnant could not be definitively ruled out.  Here, we re-examine all viable possibilities for the origin of the radio source using the new constraints provided by our higher-resolution $Ku$-band VLA observations and the ALMA CO(2--1) data.

%%%%%%%%%%%%%%%%%%%%%%%%%%%%%%%%%%%%%%%%%%%%%%%
\subsection{Star Formation}

%%%%%%%%%%%%%%%%%%%%%%%%%%%%%%%%%%%%%%%%%%%%%%%
\subsubsection{Radio Star Formation Rate}
Evidence for low-level star formation in the center of NGC\,404 includes the presence of young stars ($\lesssim$ 100~Myr) in the integrated spectra of the nuclear star cluster \citep{seth+10, bouchard+10} within the inner 0.7$^{\prime \prime}$, a central population of young stars detected in ultraviolet observations \citep{maoz+98, maoz+05}, and soft nuclear X-ray emission with a diffuse morphology \citep{binder+11}.  
Ground-based, extinction-corrected H$\alpha$ measurements suggest an upper limit to the star formation rate (SFR) of 1 $\times$ 10$^{-3}$ M$_{\odot}$~yr$^{-1}$ in the inner 1$^{\prime \prime}$ \citep{seth+10}.
At 15~GHz, the radio-SFR calibration from \citet{murphy+11} predicts that a SFR of 1.0 $\times$ 10$^{-3}$ M$_{\odot}$ yr$^{-1}$ will produce radio emission with a flux density of 0.295 mJy, which is consistent with the integrated flux density measured in our new VLA observations (Table~\ref{tab:vla}). 
% {\bf \color{magenta} Would be good to include a comparison of the nuclear Halpha flux/SFR upper limit to the central radio flux since those are at about the same spatial scale.  See the end of Nguyen+ 2017 3.2.2.}

If the extended radio source detected in our $Ku$-band observations is indeed associated with current star formation, the radio emission could in principle be produced by thermal emission from a compact H{\tt II} region.  However, the steepness of the integrated radio spectral index (see Figure~\ref{fig:radio_spectrum}) is indicative of optically-thin, non-thermal synchrotron emission.  While this effectively rules out a current star formation origin for the radio continuum emission, synchrotron emission associated with a supernova remnant produced during a recent episode of star formation could conceivably explain the radio source.  We explore this possibility in Section~\ref{sec:SNR}.

%%%%%%%%%%%%%%%%%%%%%%%%%%%%%%%%%%%%%%%%%%%%%%
\subsubsection{Molecular Gas Constraints on Star Formation}
Additional evidence for the presence of star formation in the center of NGC\,404 includes the detection of warm \citep{seth+10} and cold (e.g., \citealt{sage+89, wiklind+90, taylor+15}, this work) molecular gas from near-infrared spectroscopy and CO(1--0) imaging, respectively.  However, the near-infrared ro-vibrational H$_2$ transitions generally trace only a small fraction of the total molecular gas mass ($\sim$1~M$_{\odot}$; \citealt{seth+10}), and the previous CO(1--0) observations lacked sufficient spatial resolution to place strong constraints on the SFR in the nucleus (inner 1$^{\prime \prime}$) of NGC\,404.  The recent detection of resolved CO(2--1) emission with ALMA allows us to accurately estimate the SFR in the center of NGC\,404 for the first time.  

Using the mass of the CO(2--1) disk of $\log$($M_{\mathrm{H}2}$/M$_{\odot}$) = 5.04 and the SFR-H$_2$ mass relation from \citet{gao+04} shown below:

\begin{equation}
\frac{\mathrm{SFR}}{\mathrm{M}_{\odot}~\mathrm{yr}^{-1}} = 1.43 \times 10^{-9} \times \frac{M_{\mathrm{H}_2}}{\mathrm{M}_{\odot}}, 
\end{equation}
we find a nuclear SFR of $1.6 \times 10^{-4}$ M$_{\odot}$ yr$^{-1}$.  
We also estimate the SFR based on the H$_2$ gas surface density, $\Sigma_{\mathrm{H}2}$, using the so-called Kenicutt-Schmidt (KS) relation \citep{kennicutt+98}.  After converting to a Kroupa stellar initial mass function \citep{kroupa+01}, this SFR calibration is as follows:

\begin{equation}
\begin{split}
\log\left( \frac{\Sigma_{\mathrm{SFR}}}{\mathrm{M}_{\odot}~\mathrm{yr}^{-1}~\mathrm{kpc}^{-2}} \right) = (1.40 \pm 0.15) \\
\times \log \left( \frac{\Sigma_{\mathrm{H}_2}}{\mathrm{M}_{\odot}~\mathrm{pc}^{-2}} \right) - (3.76 \pm 0.12),
\end{split}
\end{equation}
where $\Sigma_{\mathrm{H}_2} = M_{\mathrm{H}_2}/A_{\mathrm{H}_2}$.  To measure the surface area, $A_{\mathrm{H}_2}$, we assume an annular disk geometry with an inner radius of $r_{\mathrm{inner}}=0.25$~pc and an outer radius of $r_{\mathrm{outer}}=3$~pc.  This gives $A_{\mathrm{H}_2} = \pi (r^2_{\mathrm{outer}} - r^2_{\mathrm{inner}}) = 28.1$~pc$^{2}$.  Based on our molecular gas mass measurement in the CO(2--1) disk, this yields $\Sigma_{\mathrm{SFR}} = 18.5$~M$_{\odot}$~yr$^{-1}$~kpc$^{-2}$, which corresponds to a nuclear SFR of $5.2 \times 10^{-4}$~M$_{\odot}$~yr$^{-1}$.

These molecular gas SFR estimates are factors of about 2 to 7 times lower than the SFR upper limit from the nuclear H$\alpha$ observations of \citet{seth+10} and the radio SFR based on the 15~GHz source flux density presented in this study.  This further strengthens the case against a star formation origin for the extended radio source detected in our new VLA observations.

%%%%%%%%%%%%%%%%%%%%%%%%%%%%%%%%%%%%%%%%%%%%%%
\subsection{Supernova Remnant}
\label{sec:SNR}
While current star formation cannot account for the observed properties of the spatially-resolved radio emission, we consider the possibility of non-thermal emission associated with a supernova remnant as the origin of the radio source.  Previously, \citet{nyland+12} concluded that a single supernova remnant origin for the radio emission was unlikely.  Specific evidence used to argue against the supernova remnant scenario includes the radio-X-ray-ratio (unlike NGC\,404, supernova remnants are typically relatively faint at hard X-ray wavelengths), the deconvolved size of the compact radio source on scales of parsecs, and the steep nature of the radio spectral index.  

The 15~GHz VLA data presented here have more firmly established the extent of the radio source ($\approx$17~pc) as well as the radio spectral index ($\alpha = -1.08$).  We used this information to determine the minimum synchrotron energy in Section~\ref{sec:radio_energetics} of $E_{\mathrm{min}} \approx 1.54 \times 10^{50}$~erg.  This energy is $\sim$ 10\% of the total energy budget of a type~II core-collapse supernova ($\sim$10$^{51}$~erg; \citealt{smith+98}).  Thus, a supernova remnant origin for the radio emission in the center of NGC\,404 is energetically plausible, though we remind readers that synchrotron energy estimates based on the equipartition assumption are lower limits and carry a large degree of uncertainty.
%The high minimum synchrotron energy, along with the large size measurement from this study and the high radio-X-ray ratio uncharacteristic of typical radio supernova remnants discussed in \citet{nyland+12}, further argues against a supernova remnant origin for the radio source.

\begin{figure}
\includegraphics[clip=true, trim=0.5cm 0.5cm 0.5cm 0cm, width=3.4in]{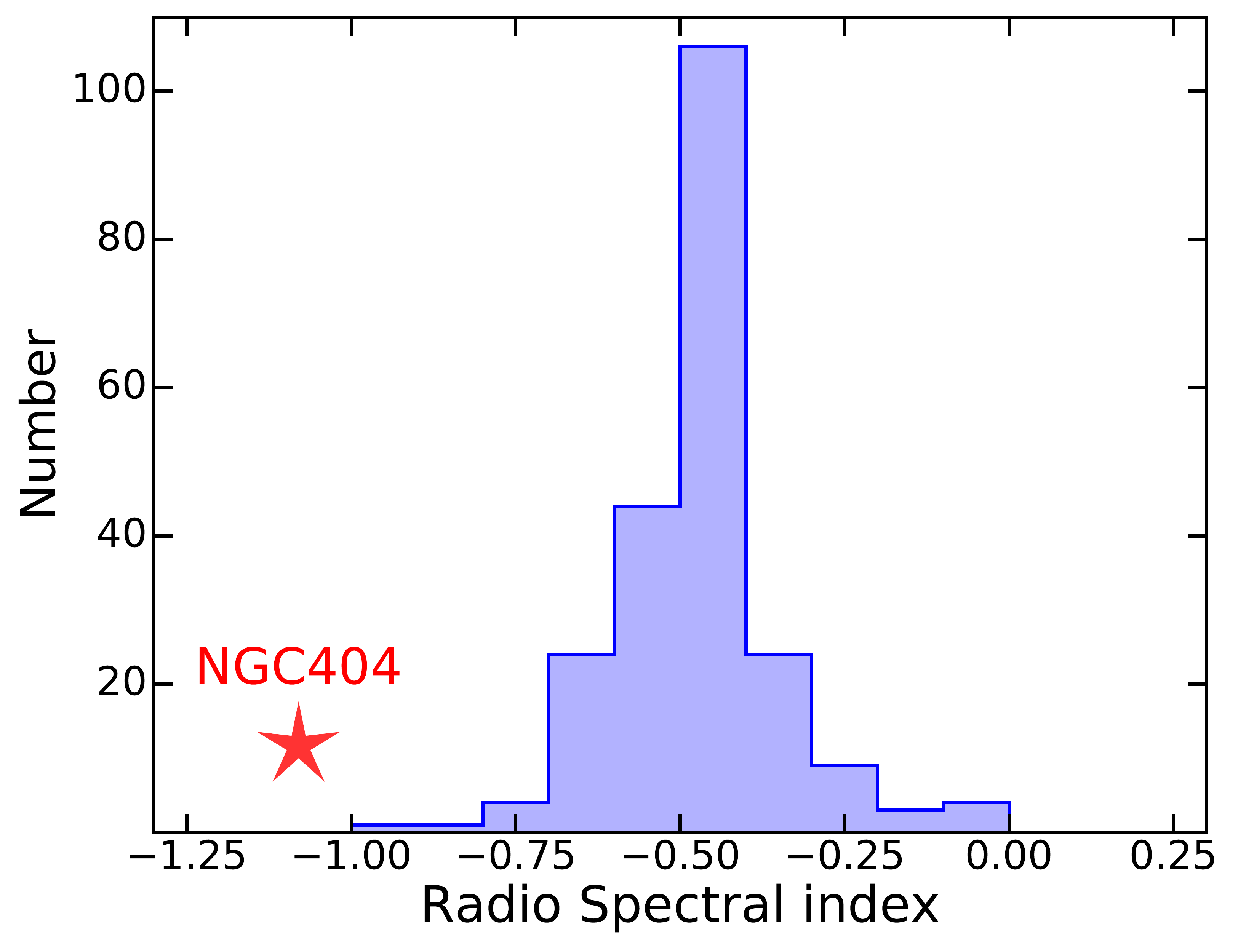} 
\caption{Distribution of Galactic radio supernova and supernova remnant spectral indices compiled in \citet{green+14}.  The radio spectral index of NGC\,404 measured from 1 to 18~GHz is marked with a red star.}
\label{fig:SNRs_alpha}
\end{figure}

We also re-examined the radio spectral index distribution of supernova remnants in light of the new, tighter constraint on the radio spectral index reported here.  Using the database of Galactic supernova remnant properties compiled by \citet{green+14}, we show the number distribution of supernova remnant spectral indices in Figure~\ref{fig:SNRs_alpha}.  NGC\,404 is shown on this figure as an obvious outlier with a considerably steeper radio spectral index compared to the Galactic supernova remnant sample.  While supernova remnants with spectral indices comparable to that of NGC\,404 are known to exist in external galaxies (e.g., \citealt{lacey+07}), these objects tend to be associated with young (i.e., less than a couple hundred years old) supernova remnants with extents on the order of one parsec or less.  The extent of NGC\,404 of 17~pc is therefore incompatible with that of a young supernova remnant characterized by a steep radio spectrum.

The radio luminosities and spectral indices of radio supernova remnants are also known to depend on the ISM properties into which the supernova remnant is expanding \citep{dubner+15}.  It is therefore possible that the radio properties of the source detected in the center of NGC\,404 are due to an extraordinary interaction between a supernova remnant and the surrounding ISM.  However, we emphasize that the circumnuclear cold molecular gas detected in our CO(2--1) ALMA observations has a more compact extent than the 15~GHz radio source (Figure~\ref{fig:radio_alma_hst}).  This suggests that supernova remnant cosmic ray re-acceleration due to a strong interaction with this phase of the ISM is unlikely to be the dominant production mechanism of the radio continuum emission.  

%%%%%%%%%%%%%%%%%%%%%%%%%%%%%%%%%%%%%%%%%%%%%%%
%%%%%%%%%%%%%%%%%%%%%%%%%%%%%%%%%%%%%%%%%%%%%%%

%%%%%%%%%%%%%%%%%%%%%%%%%%%%%%%%%%%%%%%%%%%%%%%
\subsection{Stellar Mass Black Hole}
\subsubsection{X-ray Binary}
Although the luminosity of the hard X-ray point source in the vicinity of the NGC\,404 nucleus could conceivably be produced by a single X-ray binary (XRB), previous studies have argued against this possibility.  \citet{binder+11} argued that the X-ray luminosity and power-law photon index were most consistent with a low-luminosity AGN scenario and inconsistent with the XRB interpretation.  \citet{nyland+12} reported that the high radio luminosity and high radio-X-ray-ratio were incompatible with ordinary Galactic XRBs.  They also showed that the black hole mass predicted by  the fundamental plane of black hole activity \citep{merloni+03, falcke+04} of  $\log(M_{\mathrm{BH}}) = 6.4 \pm 1.1$ M$_{\odot}$ was consistent with radio and X-ray emission produced by a MBH but not a stellar-mass black hole.  Here, our spatially-resolved map provides a robust size measurement of 17~pc, which further argues against a typical XRB origin since these objects are typically much more compact.

It is possible that a microquasar, or high-accretion-rate XRB characterized by the presence of prominent radio jets \citep{mirabel+99}, could lead to the observed radio properties of NGC\,404.  While the radio jets of microquasars are generally compact compared to the extent of NGC\,404, some microquasars are known to display particularly unusual radio properties, such as source extents of hundreds of parsecs and bright continuum emission comparable to the luminosity of the radio source in the center of NGC\,404.  The most well-known example of such a source is the Galactic microquasar SS433 \citep{fabrika+04, goodall+11}.  The extraordinary  properties of this source are the result of an interaction between the radio jets of the microquasar itself and the dense ambient medium shed by the nearby supernova remnant W50.  A handful of analogues to SS433 in external galaxies are also known, such as the microquasar S26 hosted by the nearby galaxy NGC\,7793 \citep{pakull+10, soria+10}.  The radio emission associated with objects like SS433 and S26 is believed to be produced by a jet-inflated, over-pressured bubble with an edge-brightened, cocoon-like morphology.  This type of radio continuum structure contrasts with that of NGC\,404, which consists of a central peak surrounded by less luminous, extended emission.  We therefore consider a microquasar origin for the central radio source in NGC\,404 to be unlikely.

%%%%%%%%%%%%%%%%%%%%%%%%%%%%%%%%%%%%%%%%%%%%%%%
\begin{figure*}
\includegraphics[clip=true, trim=0cm 5.25cm 0cm 5cm, width=7.2in]{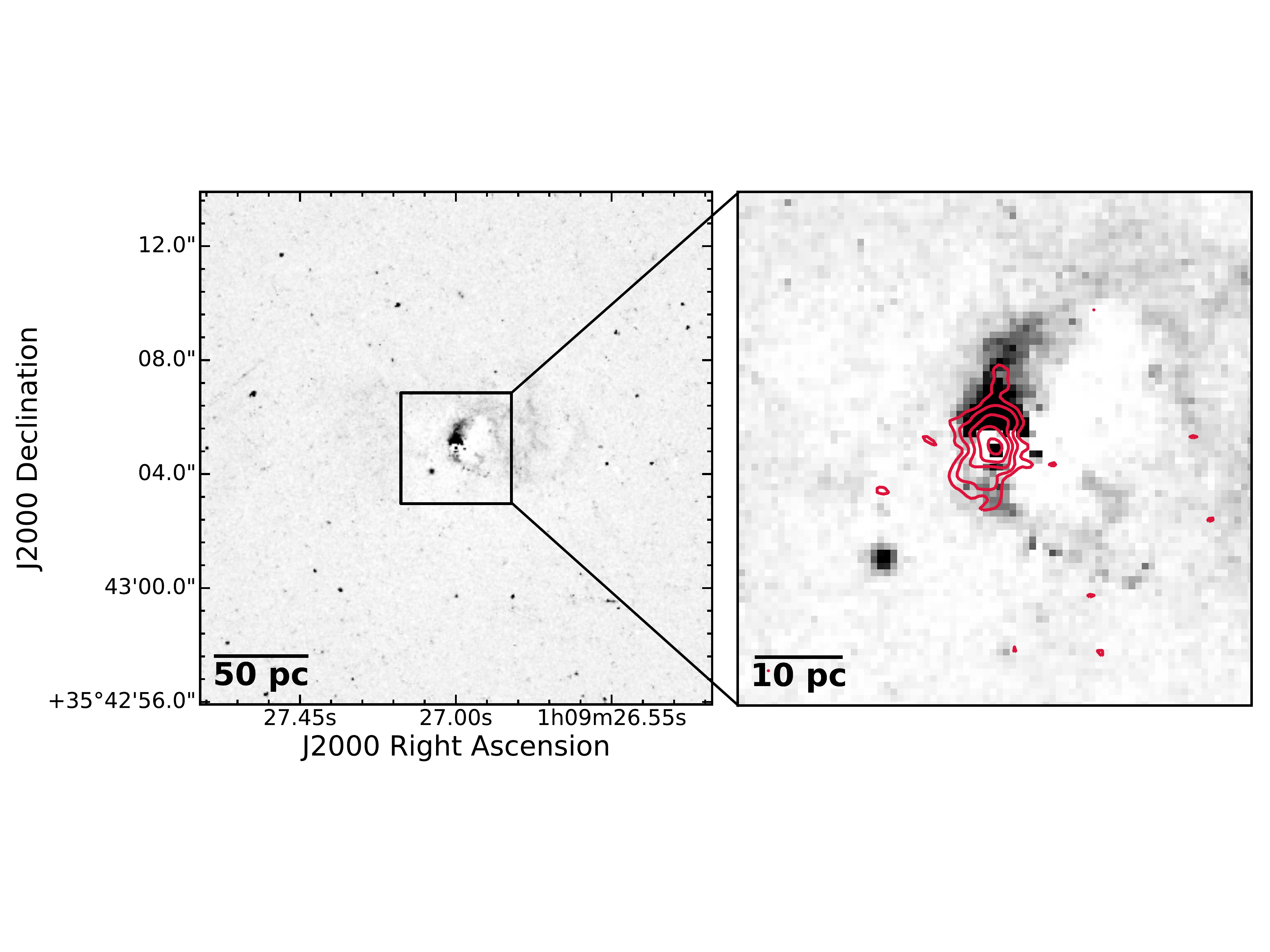} 
\caption{{\bf Left:} {\it HST} F656N$-$F814W map tracing H$\alpha$ emission shown in grayscale.  Dark regions correspond to bright H$\alpha$ emission.  The image spans 18$^{\prime \prime}$ (270~pc) per side.
{\bf Right:} Zoomed-in view of the H$\alpha$ emission in the center of NGC\,404.  The VLA 15~GHz contours are overlaid in red and are defined as in Figure~\ref{fig:radio_overlay}.  The image is 3.9$^{\prime \prime}$ (58~pc) in length per side.\\}
\label{fig:halpha}
\end{figure*}
%%%%%%%%%%%%%%%%%%%%%%%%%%%%%%%%%%%%%%%%%%%%%%%

%%%%%%%%%%%%%%%%%%%%%%%%%%%%%%%%%%%%%%%%%%%%%%%
\subsection{Accreting Massive Black Hole}
\label{sec:accreting_MBH}
\citet{nyland+12} concluded that the most likely scenario for the origin of the unresolved nuclear radio source they detected was an active nucleus powered by MBH accretion.  This conclusion was based primarily on the spatial coincidence of the radio emission with a hard X-ray point source and the optical center of the galaxy as well as the radio-X-ray ratio, defined as $R_{\mathrm{X}} = \nu L_{\nu}(5\,\mathrm{GHz})/L_{\mathrm{X}}(2-10\,\,\mathrm{keV})$ \citep{terashima+03}.  The radio-X-ray-ratio of NGC\,404 is $\log R_{\mathrm{X}} = -2.5$, consistent with the radio-loud emission characteristic of low-luminosity AGNs with radio jets or outflows \citep{nagar+05, ho+08, wrobel+12, nyland+16}.  The detection of resolved radio emission in the 15~GHz study presented here is therefore consistent with the radio-loud, low-luminosity AGN interpretation.  

%%%%%%%%%%%%%%%%%%%%%%%%%%%%%%%%%%%%%%%%%%%%%%
Although the 15~GHz emission in the nucleus of  NGC\,404 is resolved, it does not display the classic, well-collimated, core+jet morphology sometimes associated with low-luminosity AGNs on similar scales (e.g., \citealt{nagar+05}).  However, a number of studies have suggested that compact radio continuum structures lacking well-collimated morphologies may also be powered by nuclear activity (e.g., \citealt{baldi+15, mukherjee+16, soker+16}).  Explanations for the radio continuum morphology in such systems include a variable black hole accretion rate, lower bulk jet velocities, a lower black hole spin, jet precession, a higher susceptibility to interaction/entrainment with the ambient ISM, or perhaps a combination of multiple factors \citep{ulvestad+99, sikora+09, chai+12, baldi+15}.  In NGC\,404, a plausible scenario is that the radio emission originates as weak jets that quickly become confined by the ISM as they strongly interact with the surrounding gas disk, resulting in a diffuse morphology.  Such a scenario may also lead to shocks, thus contributing to the observed H$\alpha$ emission (see Section~\ref{shocks}).  The radio continuum emission could also conceivably originate from an accretion disk wind \citep{yuan+14, wong+16}.  However, we find this possibility unlikely given the steep spectral index indicative of optically-thin synchrotron emission that is incompatible with a thermal wind origin.  

Examples of nearby galaxies with radio emission associated with the central AGN that have similar radio morphologies to that seen in NGC\,404 include NGC\,1266 \citep{nyland+13} and NGC\,6764 \citep{hota+06, kharb+10}.  The radio emission in both of these galaxies is considerably more extended than that of NGC\,404 ($\sim$1~kpc vs.\ $\sim$10~pc), possibility as a result of their higher MBH masses and accretion rates.  Nevertheless, both of these objects display centrally-peaked radio emission embedded within a more diffuse, extended, and poorly-collimated structure.  In NGC\,1266, the radio emission is spatially coincident with warm ionized gas energized by shock excitation \citep{davis+12} and turbulent molecular gas \citep{alatalo+15}, suggesting the impact of the radio jet as it attempts to propagate through the ISM is responsible for the shocks.  In NGC\,6764, the diffuse morphology of the radio source is believed to result from a combination of jet precession and disruption via interaction/entrainment with the ISM \citep{kharb+10}.  While the underlying physics responsible for producing the observed radio source properties in NGC\,404 is still unclear (e.g., deceleration, ISM entrainment, jet precession, etc.), we conclude that a confined radio jet origin is a plausible scenario.

%%%%%%%%%%%%%%%%%%%%%%%%%%%%%%%%%%%%%%%%%%%%%%%
\begin{figure*}
\includegraphics[clip=true, trim=0cm 0cm 0cm 0cm, width=6.75in]{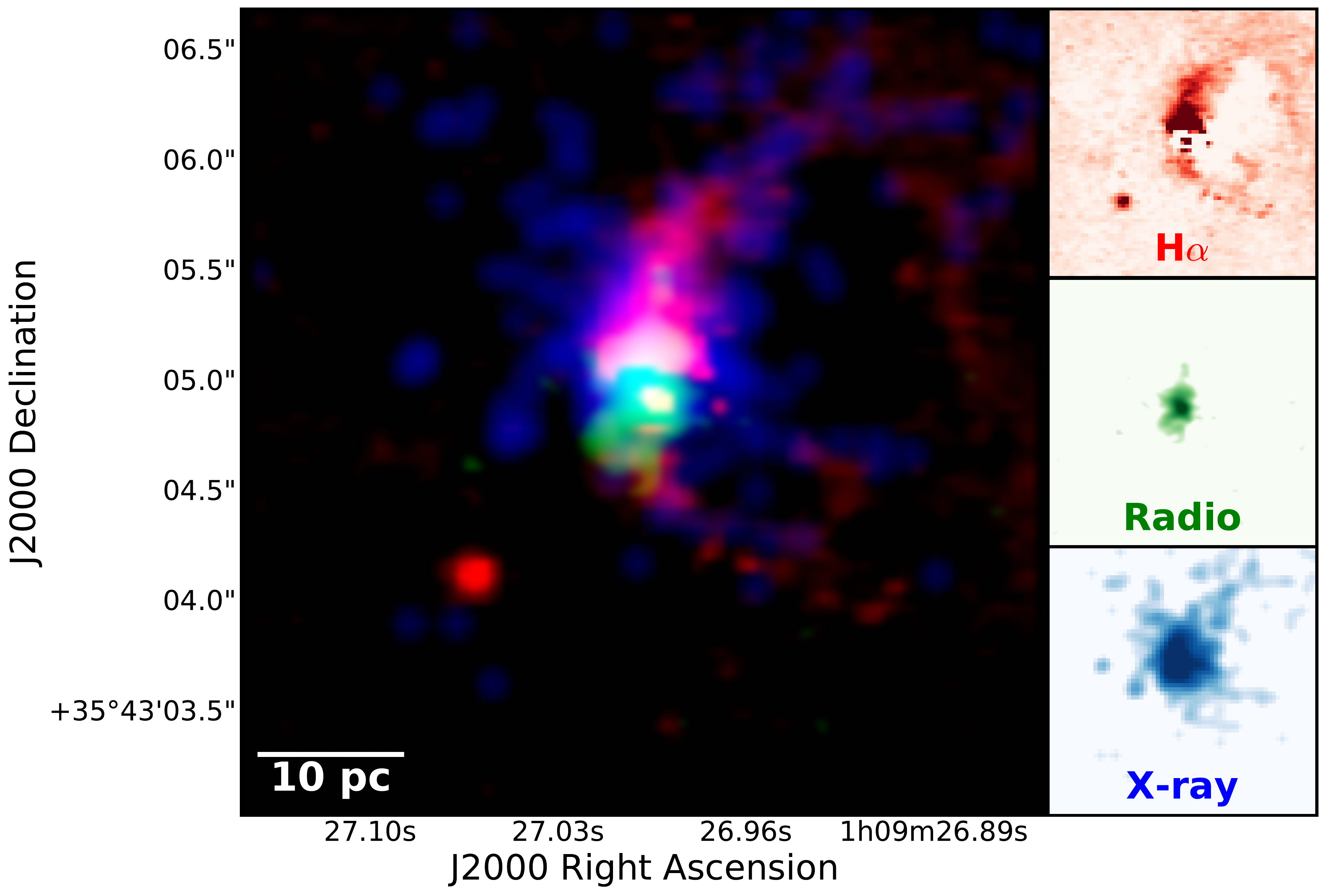} 
\caption{Composite three-color (rgb) image of the center of NGC\,404.  H$\alpha$ emission ({\it HST} F656N$-$F814W) is shown in red, 15~GHz radio continuum emission from the VLA observations presented in this work is shown in green, and soft ($0.3-2.0$ keV) X-ray emission from archival {\it Chandra} data is shown in blue.  The smaller images on the right highlight the morphology of the emission in each band.\\}
\label{fig:rgb}
\end{figure*}
%%%%%%%%%%%%%%%%%%%%%%%%%%%%%%%%%%%%%%%%%%%%%%%

%%%%%%%%%%%%%%%%%%%%%%%%%%%%%%%%%%%%%%%%%%%%%%%
%%%%%%%%%%%%%%%%%% DISCUSSION %%%%%%%%%%%%%%%%%%%%%%
%%%%%%%%%%%%%%%%%%%%%%%%%%%%%%%%%%%%%%%%%%%%%%%
%\medskip
\section{Discussion}
\label{sec:discussion}

%%%%%%%%%%%%%%%%%%%%%%%%%%%%%%%%%%%%%%%%%%%%%%%
%%%%%%%%%%%%%%%%%%%%%%%%%%%%%%%%%%%%%%%%%%%%%%%
%%%%%%%%%%%%%%%%%%%%%%%%%%%%%%%%%%%%%%%%%%%%%%%
\subsection{Shocks}
\label{shocks}
Shocks produced by supernovae, AGN-driven winds, or radio jet-ISM interactions are commonly associated with galaxy nuclei (e.g., \citealt{zakamska+14, alatalo+16, belfiore+16}), and represent a possible means of energetic feedback capable of altering the chemistry and conditions of the ISM.  We therefore consider evidence for the presence of shocks in the center of NGC\,404 from the Balmer emission line properties (including archival H$\alpha$ data from {\it HST}), soft X-ray emission from {\it Chandra}, mid-infrared data from {\it Spitzer} spectroscopy, and near-infrared spectroscopic studies from the literature.  

%%%%%%%%%%%%%%%%%%%%%%%%%%%%%%%%%%%%%%%%%%%%%%%
\subsubsection{Warm Ionized Gas}
The LINER emission line classification of NGC\,404, along with the presence of extended, nuclear H$\alpha$ emission, are consistent with the possible presence of shock excitation in the NGC\,404 nucleus as observed in other nearby galaxies (e.g., NGC\,1266; \citealt{davis+12}).  In Figure~\ref{fig:halpha}, we show an image of the {\it HST} F656-F814 emission, which provides an approximate continuum-subtracted map of the H$\alpha$ emission.  Bright, clumpy H$\alpha$ emission with an arc-like morphology is clearly visible in the circumnuclear region of NGC\,404.  This H$\alpha$ emission is brightest at the location of the northern extension of the radio source, suggesting the possibility of a link between the radio source and the production of the warm ionized gas via shock excitation.  

Previously, the extended H$\alpha$ emission in the center of NGC\,404 has been interpreted as a filament of a superbubble blown out by a circumnuclear starburst \citep{pogge+00, eracleous+02}.  However, this interpretation was based primarily on the morphology of the H$\alpha$ emission, and alternative scenarios, such as AGN feedback (e.g., NGC1052; \citealt{dopita+15}) have been advocated for other objects with similar H$\alpha$ morphologies.  We explore the possibility of ``localized'' AGN feedback in the NGC\,404 nucleus in Section~\ref{sec:fback}.

%%%%%%%%%%%%%%%%%%%%%%%%%%%%%%%%%%%%%%%%%%%%%%%
\subsubsection{Soft X-rays}
Based on previous {\it Chandra} studies, the center of NGC\,404 also hosts extended soft X-ray emission \citep{binder+11}.  In Figure~\ref{fig:rgb}, we show a three-color composite image highlighting the H$\alpha$ emission from the archival {\it HST} data, the spatially-resolved 15~GHz source from the VLA observations presented in this work, and the morphology of the soft ($0.3-2$~keV) X-ray emission from archival {\it Chandra} data.  The {\it Chandra} data have been re-binned to improve upon the default pixel resolution of 0.492$^{\prime \prime}$, thereby taking advantage of the finer positional information from spacecraft dithering and aspect corrections.  To accomplish this, we reprocessed obsid 12239 using CIAO v4.9 \citep{fruscione+06}.  We then used {\tt dmcopy} in CIAO to create a subpixel-binned image from the reprocessed event file, with image pixel scales equal to 1/8th of the native ACIS pixel scale.  Our final, improved X-ray image has also been adaptively smoothed \citep{ebeling+06} to improve the signal-to-noise ratio using the {\tt csmooth} algorithm in CIAO.  

The reprocessed soft X-ray image has allowed us to perform a more direct comparison between the {\it HST} H$\alpha$ and VLA 15~GHz data that were observed at higher spatial resolutions.  As illustrated in Figure~\ref{fig:rgb}, the re-binned soft X-ray image aligns well with the {\it HST} H$\alpha$ continuum-subtracted image, tracing the arc-like structure that is clearly visible in the warm ionized gas.  The good agreement between the warm ionized gas and soft X-rays further supports the possibility of shocks in the center of NGC\,404.  The radio continuum source also appears to spatially overlap with the diffuse X-ray and warm ionized gas emission.  The spectral index of the radio source rules out a thermal origin, suggesting that a shock origin for the 15~GHz emission is unlikely.  However, we speculate that the interaction between the non-thermal radio continuum emission and the surrounding gas could contribute to shock excitation in the vicinity of the nucleus. 

%%%%%%%%%%%%%%%%%%%%%%%%%%%%%%%%%%%%%%%%%%%%%%%
\begin{figure*}
\includegraphics[clip=true, trim=0cm 5.0cm 0cm 4.5cm, width=7.1in]{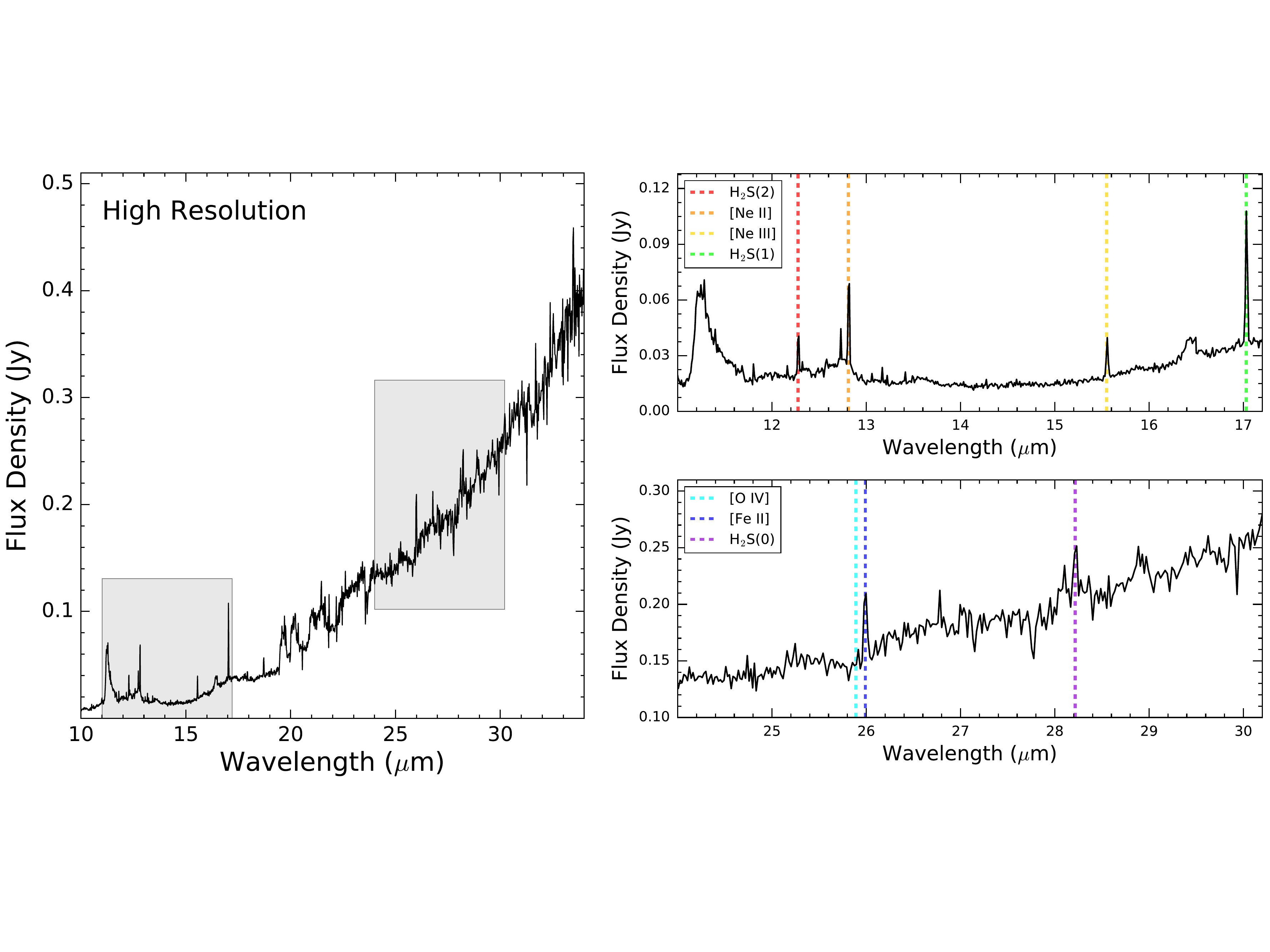} 
\caption{Archival high-resolution {\it Spitzer} IRS spectral line data obtained from CASSIS highlighting the presence of shock tracers ([Fe~{\sc ii}] and the H$_2$ lines) as well as bright mid-infrared lines commonly associated with active galaxies.  The gray-shaded regions in the main plot on the left highlight the zoomed-in views shown on the upper and lower right panels from 11 to 17.2~$\mu$m and 24 to 30.2~$\mu$m, respectively.  The prominent, unlabeled feature in the upper right plot around 11.3~$\mu$m is PAH emission (see Table~\ref{tab:spitzer2}).\\}
\label{fig:spitzer_irs}
\end{figure*}
%%%%%%%%%%%%%%%%%%%%%%%%%%%%%%%%%%%%%%%%%%%%%%%

%%%%%%%%%%%%%%%%%%%%%%%%%%%%%%%%%%%%%%%%%%%%%%%
\subsubsection{Infrared Spectroscopy}
Near-infrared spectroscopic data with sub-arcsecond spatial resolution also indicate that shocks may be present in the nucleus of NGC\,404 based on the detection of strong [Fe~{\sc ii}] and H$_2$ \citep{larkin+98, seth+10, mason+15}, tracers commonly associated with shocked gas energized by supernova remnants or radio jets (e.g., \citealt{mouri+00, alonso-herrero+97, rodriguez-ardila+04, riffel+13}).  \citet{seth+10} investigated the origin of the ro-vibrational H$_2$ emission on sub-arcsecond scales by computing the line ratio of the H$_2$ 2--1 S(1) and 1--0 S(1) lines.  They found a value of 0.11 to 0.16, which is consistent with either shocks or X-ray excitation in a dense medium ($n > 10^4$~ cm$^{-3}$).

To search for signposts of shocks in the mid-infrared, we obtained archival {\it Spitzer} Infrared Spectrograph (IRS; \citealt{houck+04}) data from The Cornell Atlas of Spitzer/IRS Sources (CASSIS\footnote{The {\it Spitzer} IRS data obtained from CASSIS correspond to AOR 10514944 and were automatically reduced using the S18.18 pipeline \citep{lebouteiller+11}.}).  We show the high-resolution IRS spectrum from the combined Short-high and Long-high modules ($10.0 \leq \lambda/\mu \mathrm{m} \leq 36.9$) in Figure~\ref{fig:spitzer_irs}.  These modules have slit sizes of 4.7$^{\prime \prime}$ $\times$ 11.3$^{\prime \prime}$ (70~pc $\times$ 168~pc) and 11.1$^{\prime \prime}$ $\times$ 22.3$^{\prime \prime}$ (165~pc $\times$ 331~pc), respectively, covering the entire NGC\,404 region of interest.  

The {\it Spitzer} IRS data have been published in \citet{pereira-santaella+10}, however, only the fine-structure lines [Ne~{\sc ii}], [Ne~{\sc iii}], and [O~{\sc iv}] were identified and measured ([Ne~{\sc v}] is not detected).  We identify these lines in Figure~\ref{fig:spitzer_irs}, along with previously unpublished detections of pure rotational transitions of molecular hydrogen and strong [Fe~{\sc ii}] emission at 26~$\mu$m.  The fluxes of the fine-structure and molecular hydrogen lines are presented in Table~\ref{tab:spitzer}.  We note that the detected fine-structure lines may arise via AGN photoionization (\citealt{veilleux+87}) or shocks (\citealt{dopita+96}), and distinguishing between these possibilities is challenging due to model degeneracies (e.g., \citealt{groves+06}).  

The rotational H$_2$ transitions detected in the {\it Spitzer} IRS data may offer additional constraints.  The presence of the warm molecular hydrogen emission is substantial given the relatively low level of star formation in the central region of this galaxy, since rotationally-excited H$_2$ lines in most galaxies are powered by FUV radiation associated with photodissociation regions (PDRs).  AGN excitation by photoionization or X-ray heating may also lead to the production of mid-infrared H$_2$ lines, though given the low hard X-ray luminosity of the NGC\,404 nucleus we find such a scenario unlikely \citep{lanz+15} (although we cannot rule out a contribution by lower-energy X-ray photons associated with supernova remnants).

In extreme cases, rotational H$_2$ lines can also be produced in shocks.  Previous studies have argued that prominent H$_2$ emission associated with radio galaxies and less powerful low-luminosity AGNs is predominantly produced via shock excitation \citep{roussel+07, ogle+10, guillard+12, alatalo+15}.  The shock front in Stephan's Quintet also shows significant warm H$_2$ emission \citep{appleton+06,guillard+12b}. Hickson Compact Group galaxies undergoing rapid transformation are another class of objects that show elevated warm H$_2$ emission, originating from shocks in the systems \citep{cluver+13}. The presence of the rotationally-excited H$_2$ lines and the 26~$\mu$m [Fe\,{\sc ii}] feature in the mid-infrared, as well as the near-infrared shock tracers H$_2$ and [Fe\,{\sc ii}] in the $K$ band, suggest that shocks may be present in the center of NGC\,404, and may even be a dominant excitation mechanism for the gas. To test this possibility, we compare tracers able to differentiate between dominant excitation mechanisms: the warm H$_2$ emission traced by the {\em Spitzer} IRS spectrum, and the 7.7~\micron\ polycyclic aromatic hydrocarbon (PAH) emission. 

%%%%%%%%%%%%%%%%%%%%%%%%%%%%%%%%%%%%%%%%%%%%%%%
\begin{table}
\begin{center}
\caption{Summary of Bright Mid-infrared Spectral Lines}
\label{tab:spitzer}
\begin{tabular}{ccccc}
\hline \hline
Line & $\lambda$ & Flux & Error \\ 
  & ($\mu$m) & (10$^{-17}$ W~m$^{-2}$) & (10$^{-17}$ W~m$^{-2}$) \\ 
(1) & (2) & (3) & (4) \\ 
\hline
 H$_2$ $0-0$ S(2) & 12.28 & 1.80 & 0.30 \\ 
$[\mathrm{Ne}\,${\sc ii}$]$  & 12.81 & 3.09 & 0.30 \\ 
$[\mathrm{Ne}\,${\sc iii}$]$ & 15.55 & 1.85 & 0.20 \\ 
H$_2$ $0-0$ S(1) & 17.03 & 5.60 & 0.80 \\ 
$[\mathrm{O}\,${\sc iv}$]$ & 25.89 & 0.79 & 0.20 \\ 
$[\mathrm{Fe}\,${\sc ii}$]$& 25.99 & 1.60 & 0.20 \\ 
H$_2$ $0-0$ S(0) & 28.22 & 1.42 & 0.20  \\ 
\hline \hline
\end{tabular}
\end{center}

\medskip
{{\it Notes} - The properties of the mid-infrared lines detected in archival, high-resolution {\it Spitzer} IRS data in this table were measured using IRAF.  Column (1): line name.  Column (2): central wavelength.  Column (3): line flux.  Column (4): uncertainty in the line flux measurement from Column (3).}
\end{table}
%%%%%%%%%%%%%%%%%%%%%%%%%%%%%%%%%%%%%%%%%%%%%%%

\begin{figure}
\includegraphics[clip=true, trim=0cm 0cm 0cm 0cm, width=3.7in]{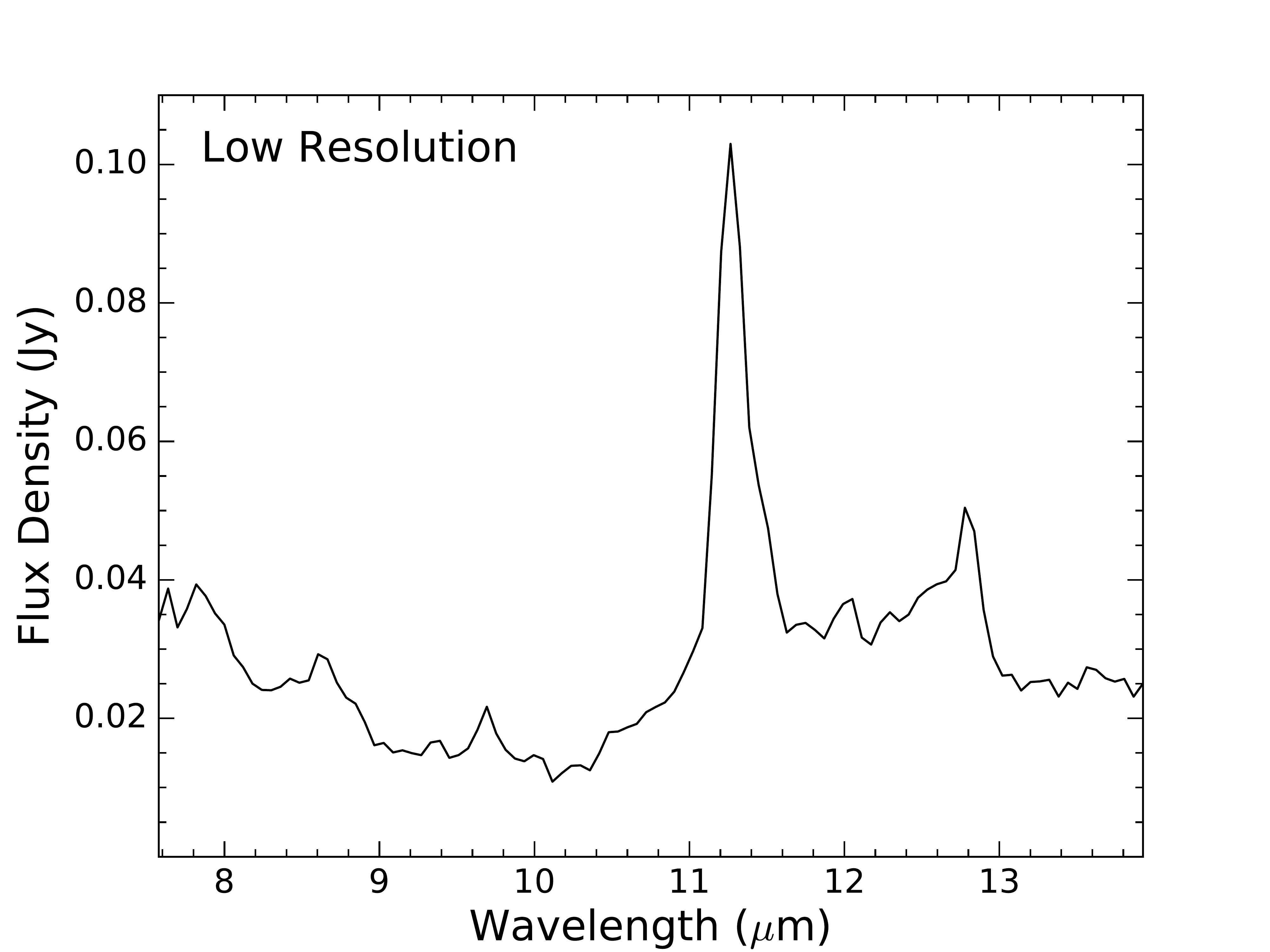} 
\caption{Archival low-resolution {\it Spitzer} IRS spectral line data obtained from CASSIS highlighting the detected PAH features listed in Table~\ref{tab:spitzer2}.}
\label{fig:spitzer_irs_low}
\end{figure}

In Figure~\ref{fig:spitzer_irs_low}, we show the low-resolution IRS spectrum, which contains the 7.7~\micron\ PAH emission region.  Data are only available from the SL1 order of the Short-low module, in the wavelength range 7.4 to 14.5 $\mu$m.  The slit size corresponding to these data is 3.7$^{\prime \prime}$ $\times$ 57$^{\prime \prime}$ (55~pc $\times$ 846~pc). Although this slit size is larger than the slit size associated with the high-resolution IRS data, the region of interest in NGC\,404 is sufficiently small that both likely encompass comparable emitting regions.  In Tables~\ref{tab:spitzer} and \ref{tab:spitzer2}, we provide summaries of the mid-infrared spectral line and PAH properties, respectively, from the archival {\it Spitzer} IRS data.  Measurements of the PAH features were obtained using the spectral fitting tool PAHFIT \citep{smith+07}.  Following \citet{ogle+10}, we calculate the ratio of the sum of the rotational H$_2$ transitions identified in Table~\ref{tab:spitzer} and the PAH feature near 7.7~$\mu$m, and find H$_2$/PAH$_{7.7\,\mu\mathrm{m}}$ $\sim 0.7$.  This ratio is well within the definition for a molecular hydrogen emitting galaxy (MOHEG) given in \citet{ogle+10} of $L(\mathrm{H}_2)/L(\mathrm{PAH}_{7.7\,\mu\mathrm{m}}) > 0.04$.  This definition encompasses a wide range of AGNs, including powerful FRI/FRII \citep{fanaroff+74} radio galaxies \citep{ogle+10}, as well as nearby low-luminosity AGNs from SINGS\footnote{{\it Spitzer} Infrared Nearby Galaxies Survey \citep{kennicutt+03}.} \citep{roussel+07}. 

Confirmation of the presence of emission line ratios consistent with shocks, as well as determination of their relative importance compared to other excitation mechanisms, will ultimately require a high-angular-resolution line ratio study of the optical and infrared emission lines that is able to differentiate between activity directly related to the AGN (such as the narrow-line region) and that which is due to shocks, but is beyond the scope of this work.  Velocity information from future two-dimensional integral-field-unit studies are also strongly motivated by the data presented here to allow a more quantitative analysis of the conditions in the center of NGC\,404, determining the direct and indirect effects of the AGN versus other influences, such as supernova remnants and the minor merger that NGC\,404 experienced $\sim$1~Gyr ago.

%%%%%%%%%%%%%%%%%%%%%%%%%%%%%%%%%%%%%%%%%%%%%%%
%%%%%%%%%%%%%%%%%%%%%%%%%%%%%%%%%%%%%%%%%%%%%%%

%%%%%%%%%%%%%%%%%%%%%%%%%%%%%%%%%%%%%%%%%%%%%%%
\begin{table}
\begin{center}
\caption{Summary of PAH Features}
\label{tab:spitzer2}
\begin{tabular}{cccccc}
\hline \hline
$\lambda$ & FWHM & EW & Flux & Error \\ 
($\mu$m) & ($\mu$m) &  ($\mu$m) & (10$^{-17}$ W~m$^{-2}$) & (10$^{-17}$ W~m$^{-2}$) \\ 
(1) & (2) & (3) & (4) & (5) \\ 
\hline
7.60& 0.334 & 0.196 & 12.20 & 3.85 \\ 
7.85 & 0.416 & 0.644 & 35.60 & 3.31 \\ 
8.61 & 0.336 & 0.616 & 24.40 & 1.72 \\ 
10.68 & 0.214 & 0.189 & 4.76 & 0.56 \\ 
11.23 & 0.135 & 0.953 & 24.80 & 0.98 \\ 
11.33 & 0.363 & 2.480 & 65.20 & 1.42 \\ 
11.99 & 0.540 & 0.932 & 27.70 & 1.47 \\ 
12.62 & 0.530 & 0.975 & 33.70 & 1.56 \\ 
13.48 & 0.539 & 0.083 & 3.62 & 1.75 \\ 
\hline \hline
\end{tabular}
\end{center}

\medskip
{{\it Notes} - The properties of the mid-infrared PAH features identified in archival, low-resolution {\it Spitzer} IRS data in this table were measured using PAHFIT \citep{smith+07}.  Column (1): central wavelength.  Column (2): full width at half maximum.  Column (3): equivalent width.  Column (4): integrated flux.  Column (5): uncertainty in the flux measurement from Column (4).}
\end{table}
%%%%%%%%%%%%%%%%%%%%%%%%%%%%%%%%%%%%%%%%%%%%%%%

%%%%%%%%%%%%%%%%%%%%%%%%%%%%%%%%%%%%%%%%%%%%%%%
%%%%%%%%%%%%%%%%%%%%%%%%%%%%%%%%%%%%%%%%%%%%%%%
%%%%%%%%%%%%%%%%%%%%%%%%%%%%%%%%%%%%%%%%%%%%%%%
\subsection{Localized MBH Feedback?}
\label{sec:fback}
Feedback from luminous AGNs may dramatically alter the cold-gas content of the ISM on galaxy-wide scales.  This form of ``global'' AGN feedback is believed to typically proceed either radiatively via the direct expulsion of gas by quasar winds, or mechanically by radio-lobe-driven heating of the intracluster medium that prevents the deposition of cooled hot halo gas onto the host galaxy.  However, mounting observational and theoretical evidence for ``localized'' feedback from much lower-luminosity AGNs, including those hosted by dwarf galaxies associated with MBH masses $<10^6$~M$_{\odot}$, has begun to transform long-held beliefs that only the most powerful AGNs residing in massive clusters or associated with quasar-level activity are capable of significantly affecting their surroundings (e.g., \citealt{alatalo+11, nyland+13, mukherjee+16, godfrey+16, silk+17}).  Thus, we consider the possibility that the accreting MBH in the center of the dwarf galaxy NGC\,404 may be injecting energy into its immediate surroundings, leading to feedback on local (i.e., tens of parsecs) scales.

Localized AGN feedback driven by the central engine of NGC\,404 may manifest itself as a nuclear outflow.  Detections of outflows launched by radio jets or winds associated with low-luminosity AGNs in nearby galaxies have become increasingly common over recent years.  Examples include NGC\,1266 \citep{alatalo+11, alatalo+15, davis+12, nyland+13}, M51 \citep{querejeta+16}, and NGC3801 \citep{hota+12}, IC5063 \citep{dasyra+12, morganti+15}, Circinus \citep{zschaechner+16}, and NGC1068 \citep{garcia-burillo+14, lamastra+16}.  However, the extent to which these outflows affect the ISMs and subsequent evolution of their host galaxies remain unclear.  If localized AGN feedback in the form of an outflow is present in the center of NGC\,404, it would provide an important comparison point to the low-luminosity AGN-driven outflows in these systems, providing constraints on the types of environments in which localized AGN feedback operates as well as the scope of the feedback.  

Although convincing evidence for a nuclear outflow is currently lacking, \citet{nguyen+17} note the presence of non-gravitational motions in the hot H$_2$ in the center of NGC\,404 that prevent them from obtaining a reliable MBH mass measurement from the gas dynamics.  One possible explanation for the observed non-gravitational motions is the presence of an outflow.  However, gas inflow, or confusion with foreground gas located at a further distance from the NGC\,404 nucleus, could also explain the H$_2$ properties.  Indirectly, evidence for the presence of shocks also suggests that AGN feedback could be present in the NGC\,404 nucleus, though as discussed in Section~\ref{shocks}, other mechanisms capable of producing shocks that are unrelated to excitation by jets/AGN winds are also possible.  

The H$\alpha$ morphology shown in Figure~\ref{fig:halpha} could be the remnant of a bubble blown by a recent starburst, ram-pressure-shocked gas fueling the central AGN, or an outflow with an ionization-cone-like structure.  Ionization cones are commonly observed in association with Seyfert nuclei (e.g., \citealt{pogge+88, muller-sanchez+11, fischer+13, maksym+16}), though examples exist in the literature of ionization cones associated with LINERs as well (e.g., \citealt{dopita+15}).  These objects are believed to be powered by AGN winds or radio jets \citep{netzer+15}.  We emphasize that the presence of an ionization cone in the center of NGC\,404, while plausible, remains speculative given the currently available data.  Future observations will ultimately be needed to confirm or rule-out the possibility of localized AGN feedback in the center of NGC\,404.  We therefore advocate for additional high-angular-resolution, multi-transition spectroscopic studies of the molecular gas in the NGC\,404 nucleus to provide constraints on the ISM conditions and excitation mechanisms.  

%%%%%%%%%%%%%%%%%%%%%%%%%%%%%%%%%%%%%%%%%%%%%%%
%%%%%%%%%%%%%%%%%%%%%%%%%%%%%%%%%%%%%%%%%%%%%%%
\subsection{Implications for Galaxy Evolution}

%%%%%%%%%%%%%%%%%%%%%%%%%%%%%%%%%%%%%%%%%%%%%%%
%%%%%%%%%%%%%%%%%%%%%%%%%%%%%%%%%%%%%%%%%%%%%%%
\subsubsection{A Dwarf Galaxy in Transition?}
Previous studies of the stellar population of NGC\,404 have revealed that the circumnuclear region is dominated by stars with ages of $\sim$1~Gyr (e.g., \citealt{nguyen+17}).  This is consistent with a large, centrally-concentrated population of A stars, a characteristic feature of post-starburst galaxies (e.g., \citealt{quintero+04, goto+07}).  Since NGC\,404 harbours prominent H$\alpha$ emission, it would be disqualified from many traditional post-starburst diagnostics that require low levels of H$\alpha$ emission to rule-out significant levels of current star formation.  Such post-starburst selection constraints aim to ensure that candidate post-starburst galaxies are likely to have experienced a substantial decline in star formation over the past 1~Gyr.  However, aside from star formation, H$\alpha$ emission may be produced via a variety of other processes, including shock excitation.  

If the central H$\alpha$ emission is primarily produced by shock excitation, NGC\,404 may represent an example of a dwarf shocked post-starburst galaxy (SPOG; \citealt{alatalo+16}), with similar properties to the prototypical SPOG NGC\,1266 \citep{alatalo+15}.  Thus, NGC\,404 may belong to a class of objects that are believed to be transitioning from actively star-forming systems to quiescent, early-type galaxies.  It is therefore of interest to consider what might have led to the post-starburst stellar population in the center of NGC\,404.  A variety of mechanisms, including AGN feedback, mergers, gas stripping in dense cluster environments, and secular evolution via the consumption of gas in a starburst event, may be responsible for transitioning the stellar populations of SPOGs.  

NGC\,404 does not reside in a cluster or group environment \citep{tikhonov+03}, and levels of AGN feedback are believed to be low (though additional studies are needed to investigate the possibility of an AGN-driven-outflow origin for the nuclear H$\alpha$ emission discussed in Section~\ref{sec:fback}).  Given these constraints, the transitioning stellar population identified in the NGC\,404 nucleus is possibly a consequence of the minor merger it experienced $\sim$1~Gyr ago (e.g., \citealt{delrio+04}) followed by the consumption of the majority of the gas deposited by the merger during a subsequent star formation episode.  We speculate that minor mergers and secular evolution may be common triggering mechanisms among field dwarf galaxies classified as SPOGs.

%%%%%%%%%%%%%%%%%%%%%%%%%%%%%%%%%%%%%%%%%%%%%%%
%%%%%%%%%%%%%%%%%%%%%%%%%%%%%%%%%%%%%%%%%%%%%%%
\subsubsection{Stunted MBH Growth in Dwarf Galaxies?}
As illustrated in Figure~16 of \citet{nguyen+17}, the stellar dynamics upper limit on the MBH mass for NGC\,404 falls below the linear scaling relation between MBH mass and bulge mass ($M_{\mathrm{BH}}$--$M_{\mathrm{sph,\ast}}$) that has been established for massive early-type galaxies \citep{kormendy+13}.  This is consistent with recent evidence that low-mass MBHs, particularly those hosted by late-type galaxies with mass measurements based on maser dynamics, do not follow the $M_{\mathrm{BH}}$--$M_{\mathrm{sph,\ast}}$ relation \citep{lasker+16, greene+16}.  One interpretation is that the $M_{\mathrm{BH}}$--$M_{\mathrm{sph,\ast}}$ relation represents an upper limit or envelope of a much broader distribution of MBH masses at a given velocity dispersion, and the tight correlation is the result of current observational limitations in resolving the gravitational sphere of influence (e.g., \citealt{batcheldor+10}).  However, \citet{gultekin+11} evaluated this possibility, and after performing a series of statistical tests, concluded that it can be ruled out for massive galaxies. 

Another possibility is that the $M_{\mathrm{BH}}$--$M_{\mathrm{sph,\ast}}$ relation is broken or bent, where the linear relation for massive galaxies (i.e., those with $M_{\mathrm{BH}} \gtrsim 10^8$~M$_{\odot}$) steepens to a nearly quadratic value for galaxies hosting less massive MBHs with masses in the range of $10^5 < M_{\mathrm{BH}}/\mathrm{M}_{\odot} \lesssim 10^7$ (e.g., \citealt{dubois+12, graham+15, fontanot+15, mezcua+16}).  Following analyses of the MBH and bulge masses of RGG\,118, which has the lowest-mass nuclear MBH known\footnote{The mass of the MBH in the center of RGG\,118 (also known as LEDA\,87300 and PGC\,87300) is $\approx$50,000~M$_{\odot}$ and was measured using the virial MBH mass measurement technique \citep{greene+05}.} \citep{baldassare+15,graham+16}, this steeper $M_{\mathrm{BH}}$--$M_{\mathrm{sph,\ast}}$ may also hold at even lower MBH masses extending to the intermediate-mass black hole regime ($10^2 < M_{\mathrm{BH}}/\mathrm{M}_{\odot} < 10^5$).

The bending of the $M_{\mathrm{BH}}$--$M_{\mathrm{sph,\ast}}$ relation suggests different evolutionary paths for high- and low-mass galaxies.  This is supported by recent statistically-complete surveys of early-type galaxies (e.g., the A{\small TLAS}$^{\rm 3D}$ survey; \citealt{cappellari+11}) that have found that massive, slowly-rotating early-types are built-up through a series of dry minor mergers, whereas less massive, more rapidly rotating early-types formed via gas-rich minor and major mergers \citep{cappellari+13}.  Thus, \citet{graham+15} suggest that for lower-mass galaxies, the bulge and MBH growth is fueled by the same reservoir of cold gas, whereas in more massive galaxies (that presumably host more massive central MBHs), AGN feedback regulates the availability of cold gas and subsequently reduces the growth of the stellar bulge.  This would naturally lead to a flatter $M_{\mathrm{BH}}$--$M_{\mathrm{sph,\ast}}$ relation at low masses.  Another explanation for the bent $M_{\mathrm{BH}}$--$M_{\mathrm{sph,\ast}}$ advocated by \citet{dubois+15} is that MBHs hosted by lower-mass galaxies are ``starved" of fuel as a result of supernova feedback that removes cold gas from the nucleus, leading to lower $M_{\mathrm{BH}}$/$M_{\mathrm{sph,\ast}}$ ratios.  This effect would decrease in significance for more massive bulges as the deeper gravitational potential makes the removal of gas via supernova feedback less efficient.

A gas-rich merger history for NGC\,404 is supported by previous studies, and evidence includes the central counter-rotating core of the NSC dominated by A stars \citep{nguyen+17} as well as the presence of an outer star-forming, cold-gas disk \citep{delrio+04, thilker+10, lelli+14}.  The recent evolutionary history of NGC\,404 is therefore consistent with the quadratic $M_{\mathrm{BH}}$--$M_{\mathrm{sph,\ast}}$ relation.  In addition to unregulated growth of the stellar bulge by sufficiently powerful AGN feedback, supernova-driven feedback may further hinder MBH accretion onto the NGC\,404 nucleus, thereby stunting its growth relative to that of the stellar bulge.  We also speculate that the putative AGN-driven radio outflow itself may further impede fueling and subsequent growth of the central MBH, as suggested for other systems (e.g., NGC3115; \citealt{wrobel+12}).

%%%%%%%%%%%%%%%%%%%%%%%%%%%%%%%%%%%%%%%%%%%%%%%
%%%%%%%%%%%%%%%%%%% SUMMARY %%%%%%%%%%%%%%%%%%%%%%
%%%%%%%%%%%%%%%%%%%%%%%%%%%%%%%%%%%%%%%%%%%%%%%	
\section{Summary and Future Work}
\label{conclu}
We have presented new VLA 12$-$18~GHz radio continuum and ALMA CO(2--1) observations of the nucleus of the low-mass MBH ($M_{\mathrm{BH}} < 1.5 \times 10^5$~M$_{\odot}$) in the center of the dwarf galaxy NGC\,404.  For the first time, we have spatially resolved the structure of the radio continuum source previously detected in the NGC\,404 nucleus at lower spatial resolution (\citealt{nyland+12}), thus highlighting the importance of high-angular-resolution, interferometric radio observations in studies of low-mass MBHs accreting at extremely low Eddington ratios.  The radio source is centrally peaked, has an extent of 1.13$^{\prime \prime}$ (17~pc), and an integrated 15~GHz luminosity of log($L$/W~Hz$^{-1}$) = 17.53.  By incorporating all radio continuum measurements of the central regions of NGC\,404 available from 1 to 18 GHz, we find a steep radio spectral index of $\alpha_{1\,\mathrm{GHz}}^{18\,\mathrm{GHz}} = -1.08 \pm 0.03$ that is consistent with optically-thin synchrotron emission.  The ALMA molecular gas data reveal a resolved, bright, circumnuclear disk as well as a clumpy, extended structure possibly related to a circumnuclear spiral arm or a disrupted GMC.  The center of the CO(2--1) disk is spatially coincident with the peak of the extended radio source, the dynamical center of NGC\,404, and the hard X-ray source identified by \citet{binder+11}, and the disk appears to be regularly rotating.  

The radio source and molecular disk are also coincident with bright, extended H$\alpha$ emission detected in narrow-band {\it HST} images.  Based on supporting evidence from the presence of extended soft X-ray emission in archival {\it Chandra} data, as well as common shock tracers detected in near- and mid-infrared spectroscopic data, the H$\alpha$ emission may be produced via shock excitation.  We further investigated the possible presence of shocks in the central region of NGC\,404 by analyzing archival {\it Spitzer} IRS data and identifying, for the first time, strong [Fe\,{\sc ii}] at 26~$\mu$m and rotational H$_2$ emission in the S(0), S(1), and S(2) transitions.  The ratio between the total warm molecular hydrogen emission and the 7.7~$\mu$m PAH feature is $\sim$0.7, indicating that mid-infrared spectrum of NGC\,404 is consistent with the MOHEG class and strengthening the case for significant shock excitation in the center of NGC\,404.  This could be a signpost of low-level, localized AGN feedback from the dwarf AGN.  However, constraints from future kinematic studies on the presence of an outflow driven by this feedback, as well as an analysis of chemical models to determine the ISM conditions in the vicinity of the nucleus of NGC\,404, are needed to confirm this intriguing possibility.

The spatial coincidence of the radio source, hard X-ray source, dynamical center of NGC\,404, and the rotating circumnuclear CO(2--1) disk point to a confined radio jet/outflow launched by the central low-mass, low-Eddington ratio MBH as the most likely origin for the radio source.  This radio outflow may also be interacting with the ISM within the inner $\sim$8~pc of the nucleus, producing shocked emission in the process.  A consequence of the low-level radio outflow may be to help maintain the low mass accretion rate of the MBH, thus contributing to the low $M_{\mathrm{BH}}$/$M_{\mathrm{sph,\ast}}$ ratio that characterizes NGC\,404 and other lower-mass galaxy spheroids.  While it is not possible to definitively rule-out an unusually luminous supernova remnant origin for the radio emission, the radio spectral index, extent of the radio source, and its centrally-peaked morphology, pose substantial challenges to the supernova remnant scenario.  This work demonstrates the importance of multi-wavelength observations for identifying the AGN signatures of a low-mass AGN, and we emphasize that similarly detailed observations will be needed to identify more objects like NGC\,404, as well as growing MBHs in high-redshift dwarf galaxies.

Although we have made progress in our understanding of the complex nature of the central engine of NGC\,404 in this study, much additional work is warranted.  Deep, interferometric radio continuum observations with slightly higher angular resolution would provide an important constraint on the presence of a central, flat-spectrum radio core embedded in the extended source identified in this work.  In the optical and infrared regimes, future, high-angular-resolution, spectroscopic analysis of the excitation mechanisms in the nucleus of NGC\,404 will be important for verifying the presence of shocks and their energetic importance compared to other excitation mechanisms.  

Additional high-resolution observations with ALMA probing the excitation mechanisms of molecular gas species such as HCO$^{+}$, HCN, and CS will help distinguish between AGN excitation and excitation driven by star formation \citep{izumi+16}, and multi-transition observations of molecular shock tracers such as SiO and HNCO (e.g., \citealt{kelly+17}, and references therein) will provide new insights on the physical conditions in the center of NGC\,404.  Finally, the high angular resolution and sensitivity of upcoming facilities such as the {\it James Webb Space Telescope} will make it possible to isolate the emission from low-luminosity AGNs residing in nearby lower-mass or dwarf galaxies from the emission of the host, which will allow the nuclear infrared SEDs and emission lines of objects like NGC\,404 to be studied in detail for the first time.  Ultimately, future statistical studies of the population of nearby low-mass AGNs will be necessary to improve our understanding of the link between MBH growth and galaxy evolution.  

%%%%%%%%%%%%%%%%%%%%%%%%%%%%%%%%%%%%%%%%%%%%%%
%%%%%%%%%%%%%%% ACKNOWLEDGMENTS %%%%%%%%%%%%%%%%%%%
%%%%%%%%%%%%%%%%%%%%%%%%%%%%%%%%%%%%%%%%%%%%%%
\section*{Acknowledgments}
We thank Preshanth Jagannathan for helpful discussions on imaging the VLA $Ku$-band data and Kyle Penner for assisting us with analyzing the archival {\it Spitzer} IRS data.  The National Radio Astronomy Observatory is a facility of the National Science Foundation operated under cooperative agreement by Associated Universities, Inc.  This paper makes use of the following ALMA data: ADS/JAO.ALMA\#2015.1.00597.S.  ALMA is a partnership of ESO (representing its member states), NSF (USA) and NINS (Japan), together with NRC (Canada) and NSC and ASIAA (Taiwan) and KASI (Republic of Korea), in cooperation with the Republic of Chile. The Joint ALMA Observatory is operated by ESO, AUI/NRAO and NAOJ.  We also utilized data from The Cornell Atlas of Spitzer/IRS Sources (CASSIS).  CASSIS is a product of the Infrared Science Center at Cornell University, supported by NASA and JPL.  This research has made use of the NASA/IPAC Extragalactic Database (NED) which is operated by the Jet Propulsion Laboratory, California Institute of Technology, under contract with the National Aeronautics and Space Administration.  
TAD acknowledges support from a Science and Technology Facilities Council Ernest Rutherford Fellowship.  Support for K.A. is provided by NASA through Hubble Fellowship grant \hbox{\#HST-HF2-51352.001} awarded by the Space Telescope Science Institute, which is operated by the Association of Universities for Research in Astronomy, Inc., for NASA, under contract NAS5-26555.  Finally, we thank the anonymous referee for helpful comments that have improved the clarity of this paper.  
\vspace{5mm}
\facilities{VLA, ALMA, {\it HST}, {\it Chandra}, {\it Spitzer}:IRS}

%%%%%%%%%%%%%%%%%%%%%%%%%%%%%%%%%%%%%%%%%%%%%%%
%%%%%%%%%%%%%%%%%% BIBLIOGRAPHY %%%%%%%%%%%%%%%%%%%%
%%%%%%%%%%%%%%%%%%%%%%%%%%%%%%%%%%%%%%%%%%%%%%%
\bibliographystyle{apj}
\bibliography{NGC404_ku_band_v9_archive}

\end{document}